\documentclass[runningheads]{llncs}
\usepackage{graphicx}
\usepackage{amsmath}
\usepackage{amssymb}
\usepackage{bbding}
\usepackage{hyperref}

\hypersetup{
	pdfauthor={Pegoraro Marco, Uysal Merih Seran, Huelsmann Tom-Hendrik, van der Aalst Wil M.P.},
	pdftitle={{Resolving Uncertain Case Identifiers in Interaction Logs: A User Study}},
	pdfsubject={Process Mining over Uncertain Data},
	pdfkeywords={Process Mining, Uncertain Event Data, Event-Case Correlation, Case Notion Discovery, Unlabeled Event Logs, Machine Learning, Neural Networks, word2vec, UI Design, UX Design},
	pdfproducer={LaTeX},
	pdfcreator={pdfLaTeX},
	bookmarksopen=true
}

\begin{document}

\title{Resolving Uncertain Case Identifiers in Interaction Logs: A User Study\thanks{We thank the Alexander von Humboldt (AvH) Stiftung for supporting our research interactions.
}}

\author{Marco Pegoraro\Envelope\orcidID{0000-0002-8997-7517} \and
	Merih Seran Uysal\orcidID{0000-0003-1115-6601} \and
	Tom-Hendrik H\"ulsmann\orcidID{0000-0001-8389-5521} \and
	Wil M.P. van der Aalst\orcidID{0000-0002-0955-6940}}

\authorrunning{M. Pegoraro et al.}
\titlerunning{Resolving Uncertain Case Identifiers in Interaction Logs: a User Study}

\institute{Chair of Process and Data Science (PADS), Department of Computer Science,\\RWTH Aachen, Aachen, Germany\\
	\email{\{pegoraro,uysal,wvdaalst\}@pads.rwth-aachen.de}\\
	\email{tom.huelsmann@rwth-aachen.de}}

\maketitle

\begin{abstract}
Modern software systems are able to record vast amounts of user actions, stored for later analysis. One of the main types of such user interaction data is click data: the digital trace of the actions of a user through the graphical elements of an application, website or software. While readily available, click data is often missing a case notion: an attribute linking events from user interactions to a specific process instance in the software. In this paper, we propose a neural network-based technique to determine a case notion for click data, thus enabling process mining and other process analysis techniques on user interaction data. We describe our method, show its scalability to datasets of large dimensions, and we validate its efficacy through a user study based on the segmented event log resulting from interaction data of a mobility sharing company. Interviews with domain experts in the company demonstrate that the case notion obtained by our method can lead to actionable process insights.

\end{abstract}

\section{Introduction}\label{sec:introduction}
In the last decades, the dramatic rise of both performance and portability of computing devices has enabled developers to design software with an ever-increasing level of sophistication. These improvements in computing performance and compactness grew in unison with their access by a larger and larger non-specialized user base, until the point of mass adoption. Such escalation in functionalities caused a subsequent increase in the complexity of software, making it more difficult to access for users. The shift from large screens of desktop computers to small displays of smartphones, tablets, and other handheld devices has strongly contributed to this increase in the intricacy of software interfaces. \emph{User interface} (UI) design and \emph{user experience} (UX) design aim to address the challenge of managing complexity, to enable users to interact easily and effectively with the software.

In designing and improving user interfaces, important sources of guidance are the records of user interaction data. While in the past enhancement to interfaces were mainly driven by manual intervention of both users of the system and designers, through survey and direct issue reporting in specialized environments, automation in all digital systems have enabled systematic and structured data collection. Many websites and apps track the actions of users, such as pageviews, clicks, and searches. Such type of information is often called \emph{click data}, of which an example is given in Table~\ref{table:data}. Click data is a prominent example of \emph{user interaction data}, a digital trace of actions which are recorded, often in real-time, when a user interacts with a system. These can then be analyzed to identify parts of the interface which need to be simplified, through, e.g., frequent itemsets analysis, pattern mining, sequence mining~\cite{DBLP:journals/tkde/Fournier-VigerW15}, or performance measures such as time spent performing a certain action or visualizing a certain page~\cite{dhandi2016comprehensive}. However, while such techniques can provide actionable insights with respect to user interface design, they do not account for an important aspect in the system operations: the \emph{process perspective}, a description of all actions in a system contributing to reach a given objective---in the case of user interfaces, the realization of the user's goal.

\begin{table}[t]
	\centering
	\caption{A sample of click data from the user interactions with the smartphone app of a German mobility sharing company. This dataset is the basis for the qualitative evaluation of the method later presented in this paper (Section~\ref{sec:qual}).}
	\begin{tabular}{|c|c|c|c|c|}
		\hline
		\textbf{timestamp}               & \textbf{screen}       & \textbf{user}  & \textbf{team}  & \textbf{os}      \\ \hline
		2021-01-25 23:00:00.939 & \texttt{pre\_booking} & b0b00 & 2070b & iOS     \\ \hline
		2021-01-25 23:00:03.435 & \texttt{tariffs}      & b0b00 & 2070b & iOS     \\ \hline
		2021-01-25 23:00:04.683 & \texttt{menu}         & 3fc0c & 02d1f & Android \\ \hline
		2021-01-25 23:00:05.507 & \texttt{my\_bookings} & 3fc0c & 02d1f & Android \\ \hline
		$\vdots$                & $\vdots$              & $\vdots$      & $\vdots$     & $\vdots$       \\ \hline
	\end{tabular}
	\label{table:data}
\end{table}

A particularly promising sub-field of data science able to account for such perspective of user interfaces is \emph{process mining}. Process mining is a discipline that aims to understand the execution of processes in a data-centric manner, by analyzing collection of historic process executions extracted by information systems---known as \emph{event logs}. Process mining techniques may be used to obtain a model of the process, to measure its conformance with normative behavior, or to analyze the performance of process instances with respect to time and costs. Data from process executions is usually represented as sorted sequences of \emph{events}, each of which is associated with an instance of the process---a \emph{case}. Although the origins of process mining are rooted in the analysis of business process data, in recent years the discipline has been successfully applied to many other contexts, with the goals of obtaining trustworthy descriptive analytics, improving process compliance, increasing time performances, and decreasing costs and wastes. Some examples are logistics~\cite{DBLP:books/sp/ReinkemeyerL20}, auditing~\cite{DBLP:books/sp/22/JansE22}, production engineering~\cite{DBLP:conf/data/AalstBGPUZ20}, and healthcare~\cite{DBLP:series/sbbpm/MansAV15}.

A number of applications of process mining techniques to user interaction data exist---prominently represented by Robotic Process Automation (see Section~\ref{sec:related}). However, towards the analysis of click data with process mining, a fundamental challenge remains: the association of event data (here, user interactions) with a \emph{process case identifier}. While each interaction logged in a database is associated with a user identifier, which is read from the current active session in the software, there is a lack of an attribute to isolate events corresponding to one single utilization of the software from beginning to end. A function able to subdivide events in sets of single instances of the process, here single utilizations of a software system, is called a \emph{case notion}. Determining the case notion in an event log is a non-trivial task, and is usually a very delicate part of event data extraction from information systems~\cite{DBLP:books/sp/Aalst16}. Aggregating user interactions into cases is of crucial importance, since the case identifier---together with the label of the executed activity and the timestamp of the event---is a fundamental attribute to reconstruct a process instance as a sequence of activities, also known as \emph{control-flow perspective} of a process instance. A vast majority of the process mining techniques available require the control-flow perspective of a process to be known.

\begin{figure}[t]
	\centering
	\includegraphics[width=0.8\textwidth]{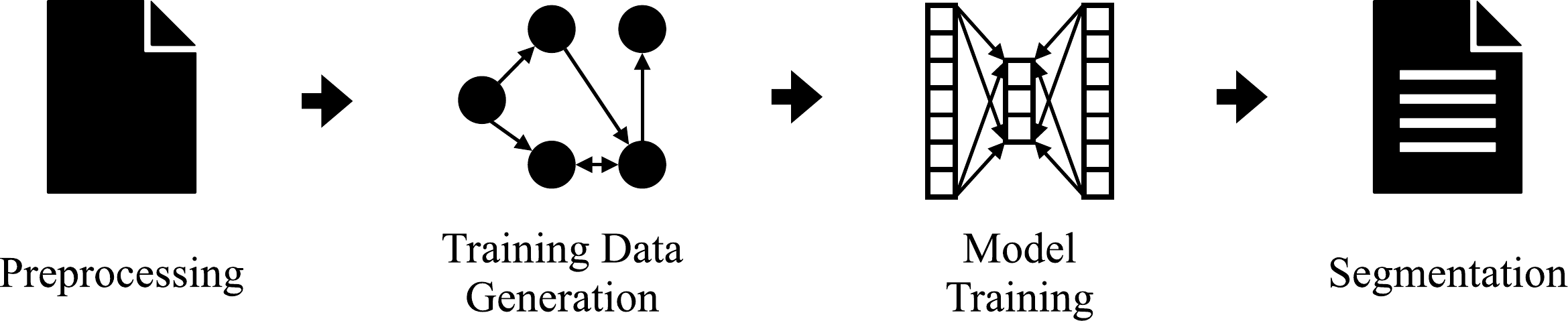}
	\caption{An overview of the different main phases of our case identifier reconstruction method.}
	\label{fig:steps}
\end{figure}

In this paper, we propose a novel case attribution approach for click data, an overview of which can be seen in Figure~\ref{fig:steps}. Our method allows us to effectively segment the sequence of interactions from a user into separate cases on the basis of normative behavior. The algorithm takes as input a collection of unsegmented user interaction and the schematic of the system in the form of a link graph, and builds a transition system able to simulate full executions of the process; then, a word2vec neural model is trained on the basis of such simulated full traces, and is then able to split an execution log into well-formed cases. We verify the effectiveness of our method by applying it to a real-life use case scenario related to a mobility sharing smartphone app. Then, we perform common process mining analyses such as process discovery on the resulting segmented log, and we conduct a user study among business owners by presenting the result of such analyses to process experts from the company. Through interviews with such experts, we assess the impact of process mining analysis techniques enabled by our event-case correlation method. Our evaluation shows that:
\begin{itemize}
	\item our method obtains a sensible case notion of an input interaction log, using comparatively weak ground truth information;
	\item our method is efficient, and is able to scale for logs of large dimensions;
	\item the resulting segmented log provides coherent, actionable insights on the process when analyzed with process mining techniques.
\end{itemize}

The remainder of the paper is organized as follows. Section~\ref{sec:preliminaries} presents preliminary concepts and constructs necessary to define our approach. Section~\ref{sec:method} illustrates a novel event-case correlation method, which allows to split a stream of interactions into cases---thus enabling process mining analyses on the resulting event log. Section~\ref{sec:perf} shows the time performance of our method at scale. Section~\ref{sec:qual} describes the results of our method on a real-life use case scenario related to a mobility sharing app, together with a discussion of interviews of process experts from the company about the impact of process mining techniques enabled by our method. Section~\ref{sec:related} examines the current literature, discussing related work and connecting our approach with existing event-case correlation methods. Finally, Section~\ref{sec:conclusion} concludes the paper.

\section{Preliminaries}\label{sec:preliminaries}

Let us start by presenting mathematical definitions for the basic structures and concepts necessary for the design of our approach.

\subsection{Process Mining}

\emph{Process mining} is a research field that lies at the intersection of established process sciences such as Business Process Management (BPM) and data science. Its goal is to extract knowledge from so-called \emph{event data} which is continuously collected during the execution of a process. A process can be any sequence of events that are carried out in order to reach a goal. Common examples include business processes such as the \emph{purchase-to-pay} process. However, in recent times, information systems have become ubiquitous and are involved in almost every aspect of modern life. Because of this omnipresence of software systems in processes, they are a prime source for event data. During their execution, such information systems produce large amounts of data in the form of logs that contain information about what actions or tasks were performed at which point in time. Process mining techniques utilize this event data in order to automatically discover new information about the underlying process. This information may then be used in order to improve the observed process in different ways. Despite its young age, the field of process mining already offers a rich ecosystem of algorithms and techniques in areas such as process discovery, conformance checking, process enhancement, and others \cite{DBLP:conf/bpm/AalstAM11,DBLP:books/sp/22/PMH2022}.

\begin{definition}[Sequence]
	Given a set $X$, a finite \emph{sequence} over $X$ of length $n$ is a function $s \in X^* : \{1, \dots, n\} \rightarrow X$, and it is written as $s = \langle s_1, s_2, \dots, s_n\rangle$. We denote with $X^*$ the set of all such sequences composed by elements of the set $X$. We denote with $\langle~\rangle$ the empty sequence, the sequence with no elements and of length 0. Over the sequence $s$ we define $|s| = n$, $s[i] = s_i$ and $x \in s \Leftrightarrow \exists_{1 \leq i \leq n} \ s = s_i$. The concatenation between two sequences is denoted with $\langle s_1, s_2, \dots, s_n\rangle \cdot \langle s'_1, s'_2, \dots, s'_m\rangle = \langle s_1, s_2, \dots, s_n, s'_1, s'_2, \dots, s'_m\rangle$. Over the sequence $\sigma$ of length $|\sigma| = n$ we define $\mathit{hd}^k(\sigma) = \langle s_1, \dots, s_{min(k, n)} \rangle$ to be the function retrieving the first $k$ elements of the sequence (if possible), and $\mathit{tl}^k(\sigma) = \langle s_{max(n-k+1,1)}, \dots, s_n \rangle$ to be the function retrieving the last $k$ elements of the sequence (if possible). Note that if $k \leq 0$ then $\mathit{hd}^k(\sigma) = \mathit{tl}^k(\sigma) = \langle~\rangle$; if $k \geq n$ then $\mathit{hd}^k(\sigma) = \mathit{tl}^k(\sigma) = \sigma$; and for all $0 \leq k \leq n$ we have that $\mathit{hd}^k(\sigma) \cdot \mathit{tl}^{n-k}(\sigma) = \sigma$.
\end{definition}

The logs containing the event data that is collected during the execution of the process are called \emph{event logs}. Event logs are a collection of individual events that at least consist of a \emph{timestamp}, the carried out \emph{activity}, and a \emph{case identifier}. These attributes represent the absolute minimum amount of information that is required for most process mining applications. Additionally, there may be other properties associated with the events, for example who carried out the activity or how long its execution did take.

\begin{definition}[Universes]
	Let the set $\mathcal{U}_I$ be the \emph{universe of event identifiers}. Let the set $\mathcal{U}_A$ be the \emph{universe of activity identifiers}. Let the set $\mathcal{U}_T$ be the totally ordered \emph{universe of timestamps}. Let the set $\mathcal{U}_U$ be the \emph{universe of users}. Let the sets $\mathcal{D}_1, \mathcal{D}_2, \dots, \mathcal{D}_n$ be the \emph{universes of attribute domains}.   
	The \emph{universe of events} is defined as $\mathcal{E} = \mathcal{U}_I \times \mathcal{U}_A \times \mathcal{U}_T \times \mathcal{U}_U \times \mathcal{D}_1 \times \mathcal{D}_2 \times \dots \times \mathcal{D}_n$.
\end{definition}

\begin{definition}[Event and Event Log]
	Any element $e \in \mathcal{E}$ is called an \emph{event}. Given an event $e = (i, a, t, u, d_1, \dots, d_n) \in \mathcal{E}$, we define the following projection functions: $\pi_I(e) = i$, $\pi_A(e) = a$, $\pi_T(e) = t$, $\pi_U(e) = u$, and $\pi_{D_j}(e) = d_j$. An \emph{event log} $L$ is a set $L \subsetneq \mathcal{E}$ where for any $e, e' \in L$, we have $\pi_I(e) = \pi_I(e') \Rightarrow e = e'$.
\end{definition}

In addition to the events themselves, a case may also be associated metadata that concerns all events of the case and can be used to further describe the underlying process instance (e.g., an order number or a customer identifier).


In order to be able to follow a single process instance throughout the process, each event is normally labeled with a case identifier, an attribute shared among all events belonging to the same process instance---a complete execution of the process to reach a certain objective, specific to each single process. Based on this, the complete event log can be grouped into multiple distinct so-called \emph{cases} that consist of sequences of events with varying lengths. The first event in a case is called the \emph{start event}, while the last event is called the \emph{end event}.

As introduced before, the existence of a timestamp, an activity, and a case identifier is generally a requirement for the majority of process mining operations. Most process mining techniques rely on the fact that a grouping of events based on the case identifier is possible. For example, consider conformance checking techniques: in order to assess if a process instance is fitting the constraints of the assumed process model, it is a requirement to be able to distinguish between the different process instances. Since this distinction is based on the case identifier, conformance checking is not possible if no such identifier is available. The same is also true for process discovery techniques, in which it is of importance to be able to identify the start and end events. In many areas of application a suitable case identifier is easily available. For example, there might be an order number, a part identifier or a distinct process id. Since these identifiers are in many cases needed during the execution of the process in order to handle the different process instances accordingly, they are generally known to the involved information systems.

However, this is not the case in all circumstances and there exists a significant number of information systems that are involved in processes, but are not process-aware. Examples of such systems include e-mail clients, that may be aware of the recipient but not the concrete case, or machines in production environments that do not have an understanding of the whole production line. In addition to that, there also exist use cases in which the definition of a case is not straightforward and it is therefore not possible to directly assign case identifiers. As introduced before, the analysis of user behavior based on recorded interaction data is an example for such a situation. A case here represents a task that the user performs. At the time of recording, it is not known when a task starts or ends. In such situations, process mining techniques cannot be applied directly to the recorded data. A preprocessing step that correlates events with cases is therefore required.

In contrast to the events in the event log, which model single events in the process, transition systems aim to encode the current state of the process and the transitions between these different states. 

\begin{definition}[Transition System]
	A \emph{transition system} is a tuple $\mathit{TS} = (S, A, T ,i,\allowbreak S^{\mathit{end}})$ where $S$ is a set of states that represent a configuration of the process, $E$ is a set consisting of the actions that can be performed in order to transition between different configurations of the system, $T \subseteq S \times A \times S$ is a set containing the transitions between configurations, $i \in S$ is the initial configuration of the process, and $S^{\mathit{end}} \subseteq S$ is the set of final configurations.
\end{definition}

Starting from the initial state $i$, the transition system can move between states according to the transition rules that are defined in $T$. A transition system can be obtained from an event log through different types of abstractions. The assumption for these abstractions is that every specific state of the process corresponds to a collection of events in the log. In general, the abstraction is either based on a window of past events, future events, or both. The size of the window is flexible and can be chosen based on the task. When there is more than a single event in the window, one has to additionally choose a representation for the events in the window. Common representations include sets, multisets and sequences of events \cite{DBLP:journals/sosym/AalstRVDKG10}. Since we will need to quantify the chances of occurring activities, we will attach probabilities to the transitions:

\begin{definition}[Probabilistic Transition System]
	A \emph{probabilistic transition system} is a tuple $\mathit{PTS} = (S, A, T ,i,\allowbreak S^{\mathit{end}})$ where $S$ is a set of states that represent a configuration of the process, $A$ is a set consisting of the activities that can be performed in order to transition between different configurations of the process, $T \colon S \times A \times S \not\to [0, 1]$ is a function expressing the probabilities of transitioning between configurations, $i \in S$ is the initial configuration of the process, and $S^{\mathit{end}} \subseteq S$ is the set of final configurations.
\end{definition}

\subsection{Embeddings}

The method presented in this paper is fundamentally based on the concept of \emph{event embeddings}~\cite{DBLP:conf/bpm/KoninckBW18}, which are themselves based on the natural language processing architecture \emph{word2vec}. The word2vec architecture allows the learning of abstract representations of words and their relations, so called \emph{embeddings}. This concept was first proposed in 2013 by Mikolov et. al. in \cite{DBLP:journals/corr/abs-1301-3781} and \cite{DBLP:conf/nips/MikolovSCCD13}. The underlying idea of word2vec is to encode the relations between words in a body of text using a shallow neural network. The resulting word embeddings are represented by vectors. The more similar the vectors of two words are according to the cosine similarity measure, the more semantically similar the words are. The technique therefore allows to capture the semantic meaning of the words, based on the way they are used in the sentences of a body of text.

\begin{figure}[h!]
	\centering
	\includegraphics[width=0.6\textwidth]{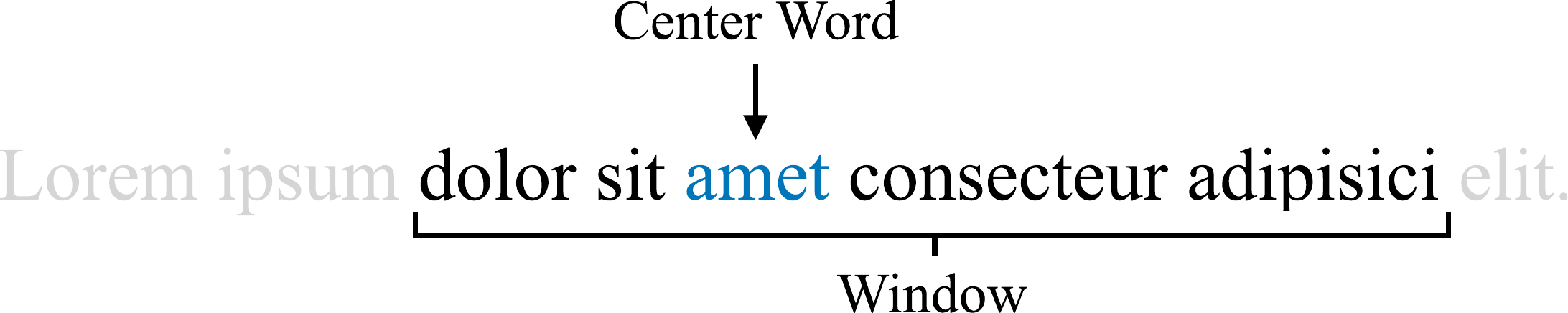}
	\caption{An example sentence from a body of text. The window has a size of five and the center word is marked in blue. The two words in front of and after the center word are the context words.}
	\label{fig:w2v_example}
\end{figure}

During the training of the two-layer neural network, a sliding window of a specified odd size is used in order to iterate over the sentences. An example for this can be found in Figure \ref{fig:w2v_example}. The word in the middle of this window is called the \emph{center word}. The words in the window before and after the center word are called \emph{context words}.

\begin{figure}[h!]
	\centering
	\includegraphics[width=0.6\textwidth]{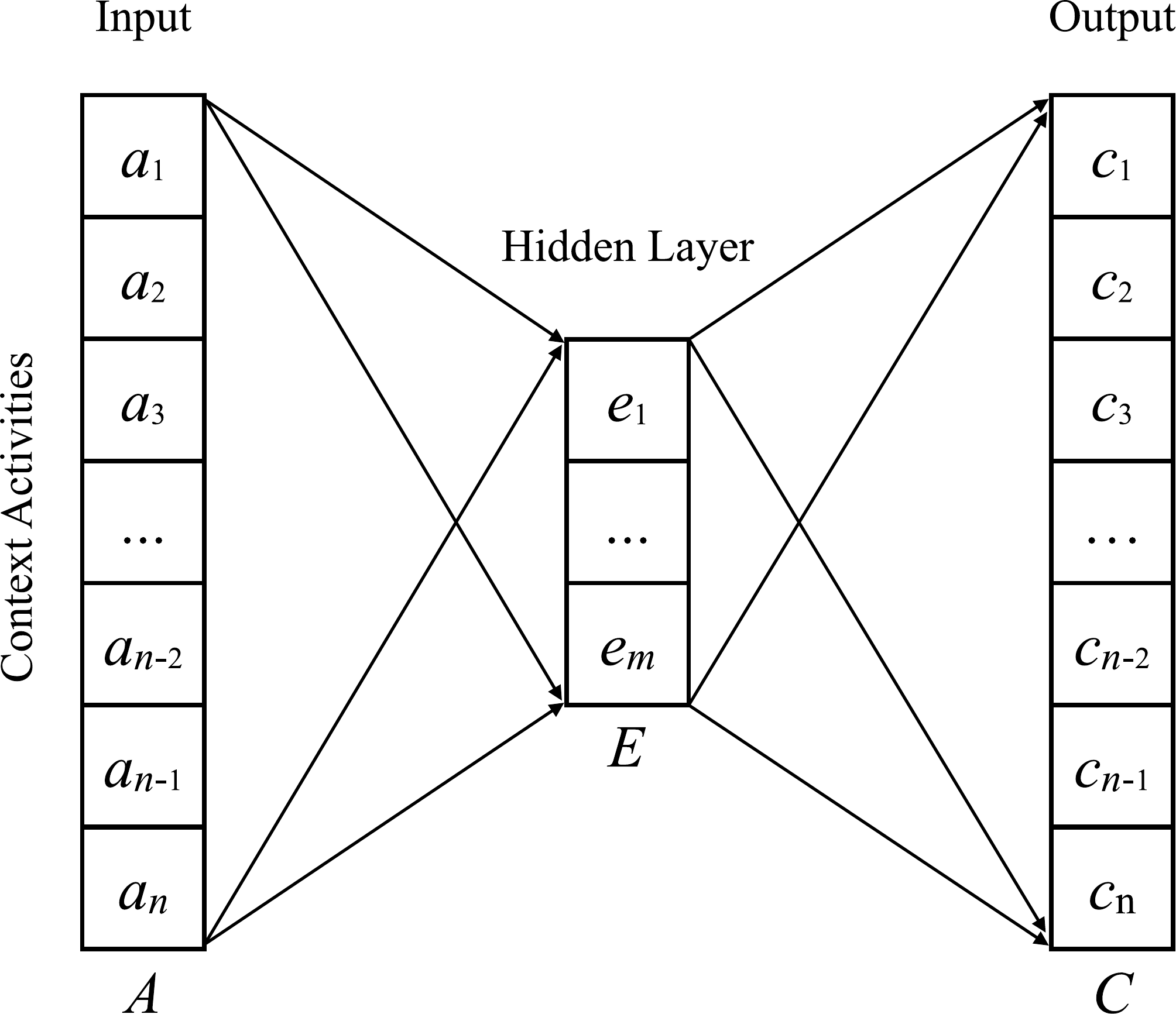}
	\caption{A graphical representation of the concept behind the event2vec architecture. The vector $A$ of size $n$ counts how often every activity occurs in the considered window. $E$ is the vector representing the event embedding of size $m$ where $m \ll n$ and vector $C$ is a one-hot encoding of the center activity in the ideal case.}
	\label{fig:e2v_diagramm}
\end{figure}

There are two different approaches to the word2vec architecture; continuous bag-of-words (CBOW) or skip-grams. The main differences between the two approaches are the input and output layers of the network. While in CBOW the frequencies of the context words are used in order to predict the center word, in the skip-gram model the center word is used to predict the context words. The order of the context words is not considered in CBOW. However, the skip-gram model does weigh the context words that are closer to the center word more heavily than those that are further away. A representation of the CBOW architecture can be found in Figure \ref{fig:e2v_diagramm}.

Both approaches produce an embedding of the context word in the form of a vector. The advantage of this architecture is that the size of the resulting embedding vectors can be freely determined through the size that is used for the hidden layer. Using this architecture, it is therefore possible to reduce the dimension of the input vector ($|V|$) considerably compared to the output embedding ($|E|$). Additionally, the word embeddings also capture information about the context in which a word is frequently used. As mentioned before, the more similar the vectors of two words, the closer the words are in meaning. In addition to this, the embeddings can also be used in order to predict the center word based on a set of given context words. Because of this versatility, the word2vec architecture is today also widely used in areas other than natural language processing, such as biology \cite{DBLP:journals/bib/Tsukiyama0FK21}, medicine \cite{10.3389/fchem.2019.00895}, or process mining \cite{DBLP:conf/smc/LakhaniN19}.

In the context of process mining, the body of text under consideration is substituted by the event log. In event embeddings, activities and traces take the role of words and sentences in word embeddings. Using this definition, the principle behind word2vec can easily be applied to event data too. Instead of the vocabulary $V$ there is the set of all possible activities $A$. During learning, each activity is associated with its embedding vector $E$, which is the output of the hidden layer. The output layer of the network $C$ ideally represents a one-hot encoding of $A$, in which only the desired center activity is mapped to one. Analogous to the word embeddings, event embeddings also capture information about the relations between the different activities. This enables the possibility to find activities that are similar to each other and allows to predict the most likely center activity based on a set of context activities. These properties of event embeddings are used by the proposed method in order to predict the boundaries between cases, by only using the sequence of activities in the interaction log. As mentioned before, this capability is not only important in the context of process mining, but also in related fields such as robotic process automation which is introduced in more detail in the next section.

\section{Method}\label{sec:method}
In this section, we illustrate our proposed method for event-case correlation on click data. As mentioned earlier, the goal is to segment the sequence of events corresponding to the interactions of every user in the database into complete process executions (cases). In fact, the click data we consider in this study have a property that we need to account for while designing our method: all events belonging to one case are contiguous in time. Thus, our goal is to determine split points for different cases in a sequence of interactions related to the same user. More concretely, if a user of the app produces the sequence of events $\langle e_1, e_2, e_3, e_4, e_5, e_6, e_7, e_8, e_9 \rangle$, our goal is to section such sequence in contiguous subsequences that represent a complete interaction---for instance, $\langle e_1, e_2, e_3, e_4 \rangle$, $\langle e_5, e_6 \rangle$, and $\langle e_7, e_8, e_9 \rangle$. Such complete interactions should reflect the behavior allowed by the system that supports the process---in the case we examine in our case and user study, such system is a mobile application. We refer to this as the \emph{log segmentation} problem, which can be considered a special case of the event-case correlation problem. In this context, ``\emph{unsegmented} log'' is synonym with ``unlabeled log''.

Rather than being based on a collection of known complete process instances as training set, the creation of our segmentation model is based on behavior described by a model of the system. A type of model particularly suited to the problem of segmentation of user interaction data---and especially click data---is the \emph{link graph}. In fact, since the activities in our process correspond to screens in the app, a graph of the links in the app is relatively easy to obtain, since it can be constructed in an automatic way by following the links between views in the software. This link graph will be the basis for our training data generation procedure.
\begin{definition}[Link Graph]
	A \emph{link graph} of a software is a graph $\mathit{LG} = (V, E)$ where $V$ is the set of pages or screens in the software, and $E \subseteq V \times V$ represents the links from a page to the next.
\end{definition}
We will use as running example the link graph of Figure~\ref{fig:dfg}. The resulting normative traces will then be used to train a neural network model based on the word2vec architecture~\cite{DBLP:conf/nips/MikolovSCCD13}, which will be able to split contiguous user interaction sequences into cases.

\subsection{Training Log Generation}\label{sec:gen}
To generate the training data, we will begin by exploiting the fact that each process case will only contain events associated with one and only one user. Let $L$ be our unsegmented log and $u \in \mathcal{U}_U$ be a user in $L$; then, we indicate with $\mathit{UI}_u$ the \emph{user interaction sequence}, a sequence of activities in a sub-log of $L$ sorted on timestamps where all events are associated with the user $u$: $\mathit{UI}_u = \langle \pi_A(e_1), \pi_A(e_2), \dots, \pi_A(e_n) \rangle$ such that $e \in \mathit{UI}_u \Rightarrow e \in L \wedge \pi_U(e) = u$, and it holds\footnote{We assume user interactions to be tied to clicks on a UI element, so no two user actions can be recorded at the same time. Thus, the total order between events is strict. Assuming a strict total ordering on events is ubiquitous in process mining.} that $\pi_T(e_1) < \pi_T(e_2) < \dots < \pi_T(e_n)$.

Our training data will be generated by simulating a transition system annotated with probabilities. Initially, for each user $u \in U$ we create a transition system $\mathit{TS}_u$ based on the sequence of user interactions $\mathit{UI}_u$. The construction of a transition system based on event data is a well-known procedure in process mining~\cite{DBLP:journals/sosym/AalstRVDKG10}, which requires to choose a state representation abstraction function $\mathit{state} \colon \mathcal{U}_A \to S_u$ and a window size (or horizon), which are process-specific. In the context of this section, we will show our method using a prefix sequence abstraction with window size 2: $\mathit{state}(s) = \mathit{tl}^2(s)$. The application of other abstraction functions is of course possible.

All such transition systems $\mathit{TS}_u$ share the same initial state $i$. To identify the end of sequences, we add a special symbol to the states $f \in S'$ to which we connect any state $s \in S$ if it appears at the end of a user interaction sequence. To traverse the transitions to the final state $f$ we utilize as placeholder the empty label $\tau$. Formally, for every user $u \in \mathcal{U}_U$ and user interaction $\mathit{UI}_u$ with length $n = |\mathit{UI}_u|$, we define $\mathit{TS}_u = (S_u, A_u, T_u, i_u, S_u^{\mathit{end}})$ as:
\begin{itemize}
	\item $S_u \,= \{\mathit{state}(\mathit{hd}^k(\mathit{UI}_u)) \mid 0 \leq k \leq n \} \cup \{f\}$
	\item $A_u = \{\mathit{UI}_u[k] \mid 0 \leq k \leq n \} \cup \{\tau\}$
	\item $T_u \,= \{(\mathit{state}(\mathit{hd}^k(\mathit{UI}_u)), \mathit{UI}_u[k + 1] ,\mathit{state}(\mathit{hd}^{k+1}(\mathit{UI}_u))) \mid 0 \leq k \leq n - 1 \} \cup \{(\mathit{state}(\mathit{hd}^n(\mathit{UI}_u)), \tau, f)\}$
	\item $i_u = \langle~\rangle$
	\item $S_u^{\mathit{end}} = \{f\}$
\end{itemize}
For instance, the user interaction $\langle M, A, B, C \rangle$ results in $S_u = \{\langle~\rangle, \langle M \rangle, \langle M, A \rangle,\allowbreak\langle A, B \rangle, \langle B, C \rangle, f \}$, $A_u = \{M, A, B, C, \tau \}$, and $T_u = \{(\langle~\rangle, M, \langle M \rangle),(\langle M \rangle, A, \langle M, A \rangle),\allowbreak(\langle M, A \rangle, B, \langle A, B \rangle), (\langle A, B \rangle, C, \langle B, C \rangle), (\langle B, C \rangle, \tau, f) \}$.

We then obtain a transition system $TS' = (S', A', T', i', S'^{\mathit{end}})$ corresponding to the entire log $L$ by merging the transition systems corresponding to the users:
\begin{itemize}
	\item $S' = \bigcup_{u \in \mathcal{U}_U} S_u$
	\item $A' = \bigcup_{u \in \mathcal{U}_U} A_u$
	\item $T' = \bigcup_{u \in \mathcal{U}_U} T_u$
	\item $i' = \langle~\rangle$
	\item $S'^{\mathit{end}} = \{f\}$
\end{itemize}
We also collect information about the frequency of each transition in the log: for the transitions $(s, a, s') = t \in T$, we define a weighting function $\omega \colon T \to \mathbb{N}$ which measures the number of occurrences of the transition $t$ throughout the entire log:
\begin{align*}
	\omega((s, a, s')) = &\left| \bigcup_{u \in \mathcal{U}_U}\{(k, u) \mid 0 \leq k \leq n - 1 \wedge  \mathit{state}(\mathit{hd}^k(\mathit{UI}_u)) = s\,\wedge\right.\\ 
	&\left.\wedge\mathit{UI}_u[k + 1] = a \wedge \mathit{state}(\mathit{hd}^{k+1}(\mathit{UI}_u)) = s' \} \vphantom{\bigcup_{u \in \mathcal{U}_U}}\right|
\end{align*}
If $t \notin T$, $\omega(t) = 0$. Through $\omega$, it is optionally possible to filter out rare behavior by deleting transitions with $\omega(t) < \epsilon$, for a small threshold $\epsilon \in \mathbb{N}$. Figure~\ref{fig:ts} shows the transition system $\mathit{TS'}$ with the chosen abstraction and window size, annotated with both frequencies and transition labels, for the user interactions $\mathit{UI}_{u_1} = \langle M, A, M, B, C \rangle$, $\mathit{UI}_{u_2} = \langle M, B, C, M \rangle$, and $\mathit{UI}_{u_3} = \langle M, A, B, C \rangle$.

In contrast to transition systems that are created based on logs that are segmented, the obtained transition system might contain states that are not reachable and transitions that are not possible according to the real process. Normally, the transition system abstraction is applied on a case-by-case basis. In our case, however, we applied the abstraction to the whole sequence of interactions that is associated with a specific user, consecutive interactions that belong to different cases will be included as undesired transitions in the transition system. In order to prune undesired transitions from the transition system, we exploit the link graph of the system: a transition in the transition system is only valid if it appears in the link graph. Unreachable states are also pruned.

\begin{figure}
	\centering
	\includegraphics[width=.25\textwidth]{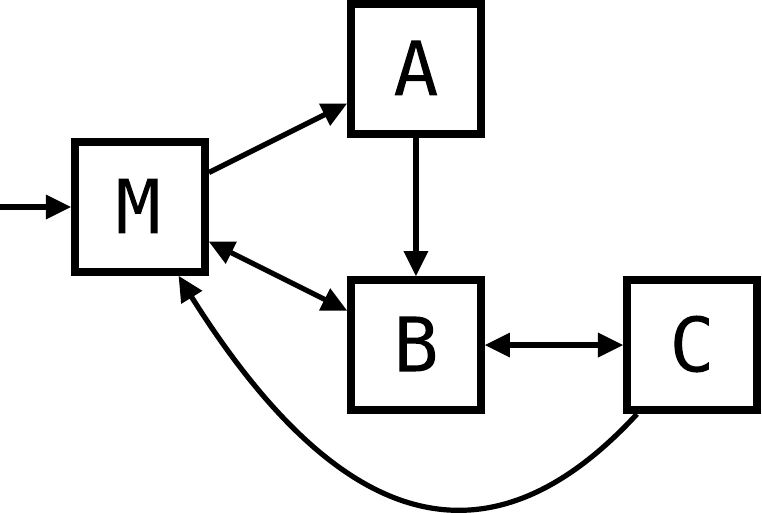}
	\caption{The link graph of a simple, fictional system that we are going to use as running example. From this process, we aim to segment the three unsegmented user interactions $\langle M, A, M, B, C \rangle$, $\langle M, B, C, M \rangle$, and $\langle M, A, B, C \rangle$.}
	\label{fig:dfg}
\end{figure}

\begin{figure}
	\centering
	\includegraphics[width=.6\textwidth]{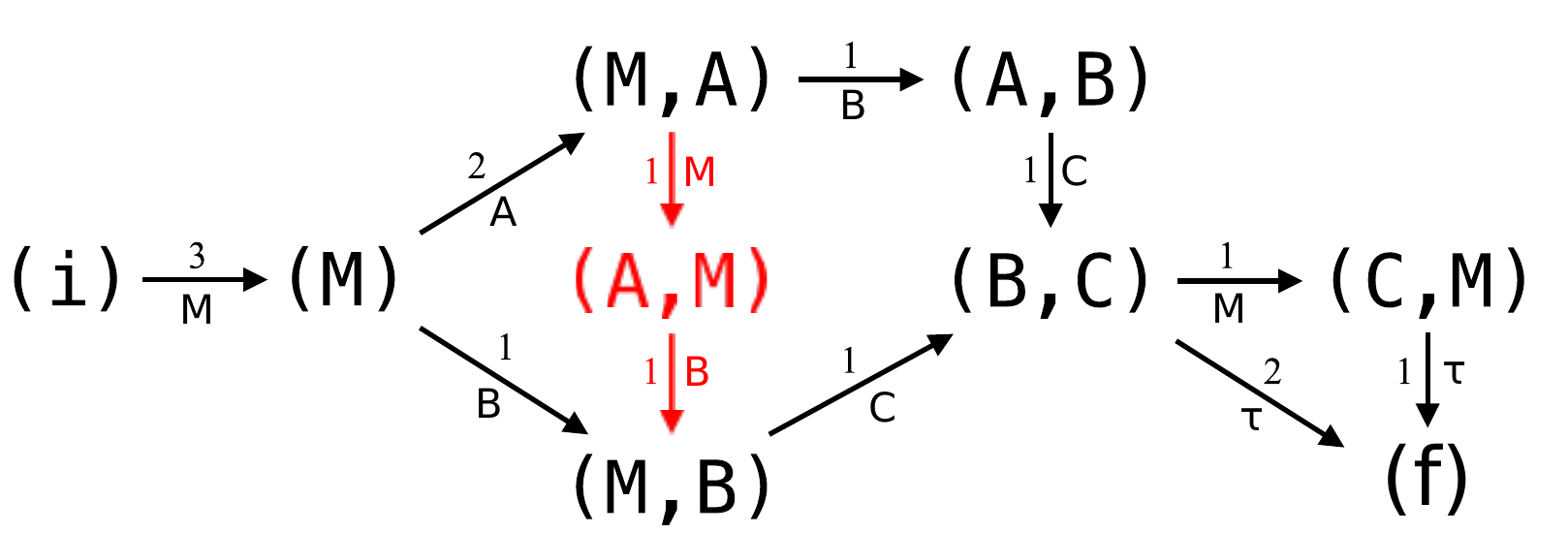}
	\caption{The transition system $\mathit{TS'}$ obtained by the user interaction data of the example (Figure~\ref{fig:dfg}). During the reduction phase, the transition $(M, A)$ to $(A, M)$ is removed, since it is not supported by the link graph ($M$ does not follow $A$). The state $(A, M)$ is not reachable and is removed entirely (in red). Consequently, the reduced transition system $\mathit{TS}_r$ is obtained.}
	\label{fig:ts}
\end{figure}

We will again assume a sequence abstraction. Given a link graph $\mathit{LG} = (V, E)$, we define the reduced transition system $\mathit{TS}_r = (S_r, A_r, T_r, i_r, S_r^{\mathit{end}})$, where:
\begin{itemize}
	\item $S_r = \bigcup_{(s, a, s') \in T} \{s, s'\}$
	\item $A_r = \{a \in \mathcal{U}_A \mid (s, a, s') \in T' \}$
	\item $T_r = \{(\langle \dots, a \rangle, a', \langle \dots, a, a' \rangle) \in T' \mid (a, a') \in E\}$
	\item $i_r = \langle~\rangle$
	\item $S_r^{\mathit{end}} = \{f\}$
\end{itemize}
Figure~\ref{fig:dfg} shows a link graph for our running example, and Figure~\ref{fig:ts} shows how this is used to reduce $\mathit{TS'}$ into $\mathit{TS}$.

Next, we define probabilities for transitions and states based on the count values for $\omega(t)$. Let $T_{\text{out}} \colon S \to \mathcal{P}(T_r)$ be  $T_{\text{out}}(s) = \{(s', a, \mathit{s''}) \in T_r \mid s' = s\}$; this function returns all outgoing transitions from a given state. The likelihood of a transition $(s, a, s') \in T_r$ is then computed with $l_{\text{trans}} \colon T_r \to [0, 1]$:

$$
l_{\text{trans}}(s, a, s') = \frac{\omega(s, a, s')}{\sum\limits_{t \in T_{\text{out}}(s)} \omega(t)}
$$

Note that if $s$ has no outgoing transition and $T_{\text{out}}(s) = \varnothing$, we have that $l_{\text{trans}}(s, a, s') = 0$ for any $a \in A$ and $s' \in S_r$. We will need two more support functions. We define $l_{\text{start}} \colon S_r \to [0, 1]$ and $l_{\text{end}} \colon S_r \to [0, 1]$ as the probabilities that a state $s \in S$ is, respectively, the initial and final state of a sequence:\\

\noindent\begin{minipage}{.5\linewidth}
	$$
	l_{\text{start}}(s) = \frac{\sum\limits_{a \in A}\omega(i, a, s)}{\sum\limits_{\substack{s' \in S_r \\ a \in A}} \omega(s', a, s)}
	$$
\end{minipage}
\begin{minipage}{.5\linewidth}
	$$
	l_{\text{end}}(s) = \frac{\omega(s, \tau, f)}{\sum\limits_{\substack{s' \in S_r \\ a \in A}} \omega(s, a, s')}
	$$
\end{minipage}\\

In our example of Figure~\ref{fig:ts}, $l_{\text{start}}((M)) = \frac{3}{3} = 1$ and $l_{\text{end}}((C, M)) = \frac{1}{3}$.

Such probability functions allow us to define the probabilistic transition system that can simulate an event log based on our dataset of user interactions. We will extend the reduced transition system $\mathit{TS}_r$ into a probabilistic transition system $\mathit{PTS} = (S, A, T, i, S^\mathit{end})$ where:
\begin{itemize}
	\item $S = S_r$
	\item $A = A_r$
	\item $T = l_{\text{trans}}$
	\item $i = i_r$
	\item $S^{\mathit{end}} = S_r^{\mathit{end}}$
\end{itemize}

Given a path of states $\langle s_1, s_2, \dots, s_n \rangle$ transitioning in $\mathit{PTS}$ through the sequence $\langle (i, a_1, s_1),\allowbreak(s_1, a_2, s_2), \dots, (s_{n - 1}, a_n, s_n), (s_n, \tau, f) \rangle$, we now have the means to compute its probability with the function $l \colon S^* \to [0, 1]$:

$$
l(\langle s_1, s_2, \dots, s_n \rangle) = l_{\text{start}}(s_1) \cdot \prod\limits_{\substack{i = 2}}^n l_{\text{trans}}(s_{i - 1}, a_i, s_{i}) \cdot l_{\text{end}}(s_n)
$$

This enables us to obtain an arbitrary number of well-formed process cases as sequences of activities $\langle a_1, a_2, \dots, a_n \rangle$, utilizing a Monte Carlo procedure. We can sample a random starting state for the case, through the probability distribution given by $l_{\text{start}}$; then, we compose a path with the probabilities provided by $l_{\text{trans}}$ and $l_{\text{end}}$. The traces sampled in this way will reflect the available user interaction data in terms of initial and final activities, and internal structure, although the procedure still allows for generalization. Such generalization is, however, controlled thanks to the pruning provided by the link graph of the system. We will refer to the set of generated traces as the training log $L_T$.

\subsection{Model Training}\label{sec:training}
The training log $L_T$ obtained in Section~\ref{sec:gen} is now used in order to train the segmentation models. The core component of the proposed method consists one or more word2vec models to detect the boundaries between cases in the input log. When applied for natural language processing, the input of a word2vec model is a corpus of sentences which consist of words. Instead of sentences built as sequences of words, we consider traces $\langle a_1, a_2, \dots, a_n \rangle$ as sequences of activities.

\begin{figure}
		\centering
		\includegraphics[width=.6\textwidth]{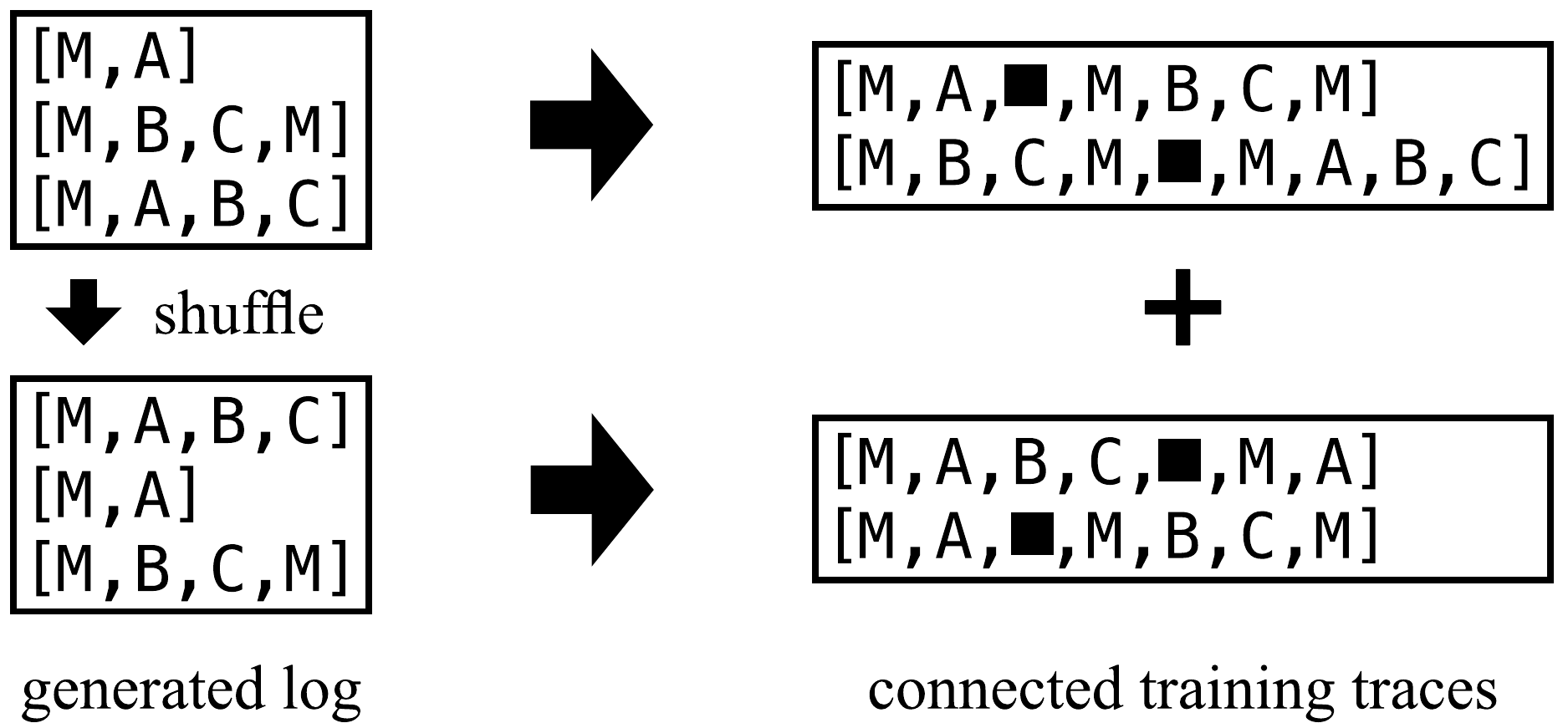}
		\caption{Construction of the training instances. Traces are shuffled and concatenated with a placeholder end activity.}
		\label{fig:training}
\end{figure}

The training log $L_T$ needs an additional processing step to be used as training set for word2vec. Given two traces $\sigma_1 \in L_T$ and $\sigma_2 \in L_T$, we build a training instance by joining them in a single sequence, concatenating them with a placeholder activity $\blacksquare$. So, for instance, the traces $\sigma_1 = \langle a_1, a_2, a_4, a_5 \rangle \in L_T$ and $\sigma_2 = \langle a_6, a_7, a_8 \rangle \in L_T$ are combined in the training sample $\langle a_1, a_2, a_4, a_5, \blacksquare, a_6, a_7, a_8 \rangle$. This is done repeatedly, shuffling the order of the traces. Figure~\ref{fig:training} shows this processing step on the running example.

\begin{figure}
		\centering
		\includegraphics[width=.6\textwidth]{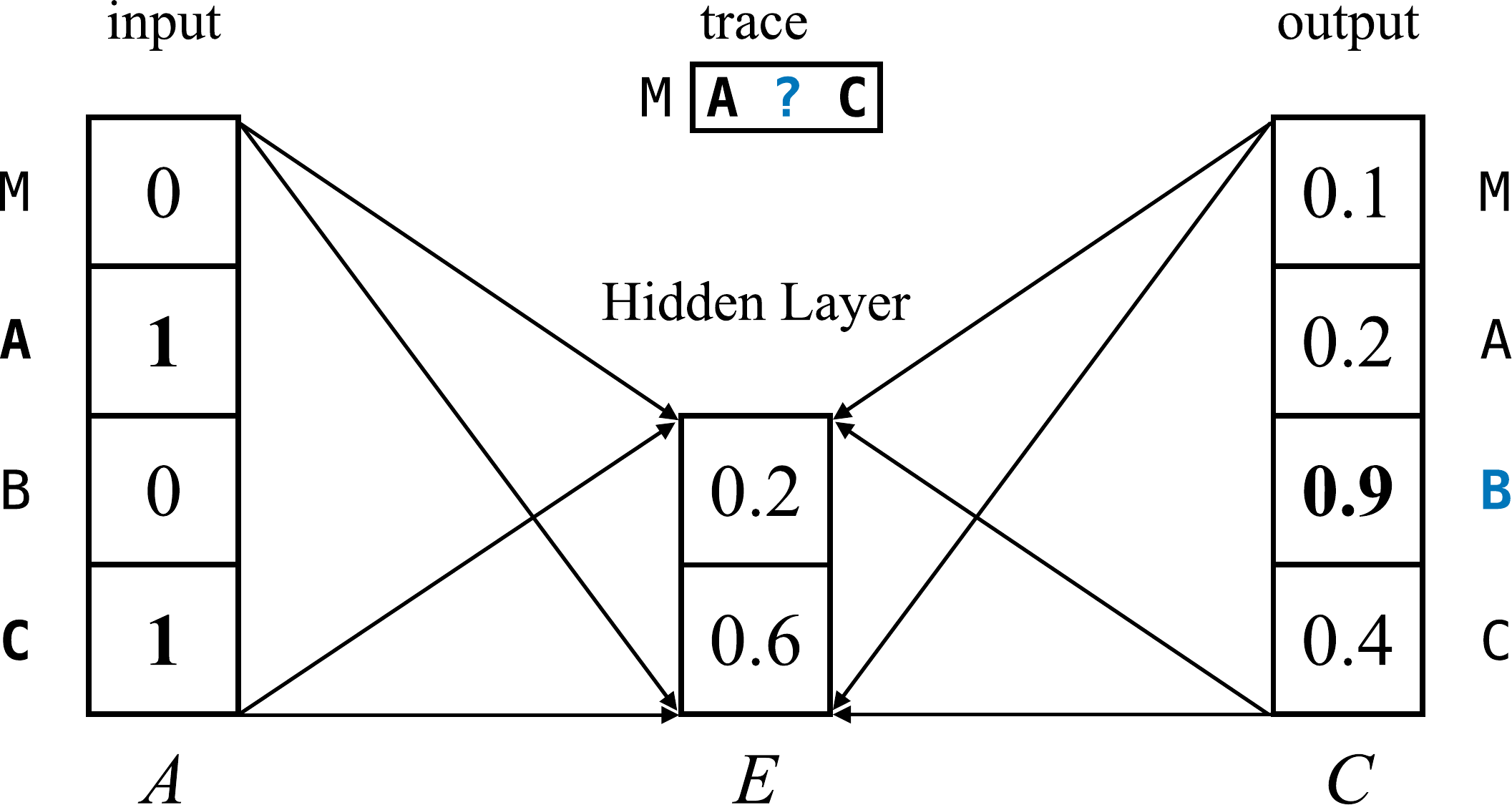}
		\caption{The word2vec neural network. Given the sequence $\langle A, ?, C \rangle$, the network produces a probability distribution over the possible activity labels for $?$.}
		\label{fig:net}
\end{figure}

The word2vec model~\cite{DBLP:conf/nips/MikolovSCCD13} consists of three layers: an input layer, a single hidden layer, and the output layer. This model has already been successfully employed in process mining to solve the problem of missing events~\cite{DBLP:conf/smc/LakhaniN19}. During training, the network reads the input sequences with a sliding window. The activity occupying the center of the sliding window is called the \emph{center action}, while the surrounding activities are called \emph{context actions}. The proposed method uses the \emph{Continuous Bag-Of-Words} (CBOW) variant of word2vec, where the context actions are introduced as input in the neural network in order to predict the center action. The error measured in the output layer is used for training in order to adjust the weights in the neural network, using the backpropagation algorithm. These forward and backward steps of the training procedure are repeated for all the positions of the sliding window and all the sequences in the training set; when fully trained, the network will output a probability distribution for the center action given the context actions. Figure~\ref{fig:net} shows an example of likelihood estimation for a center action in our running example, with a sliding window of size 3.

\subsection{Segmentation}
\begin{figure}[]
	\centering
	\includegraphics[width=.6\textwidth]{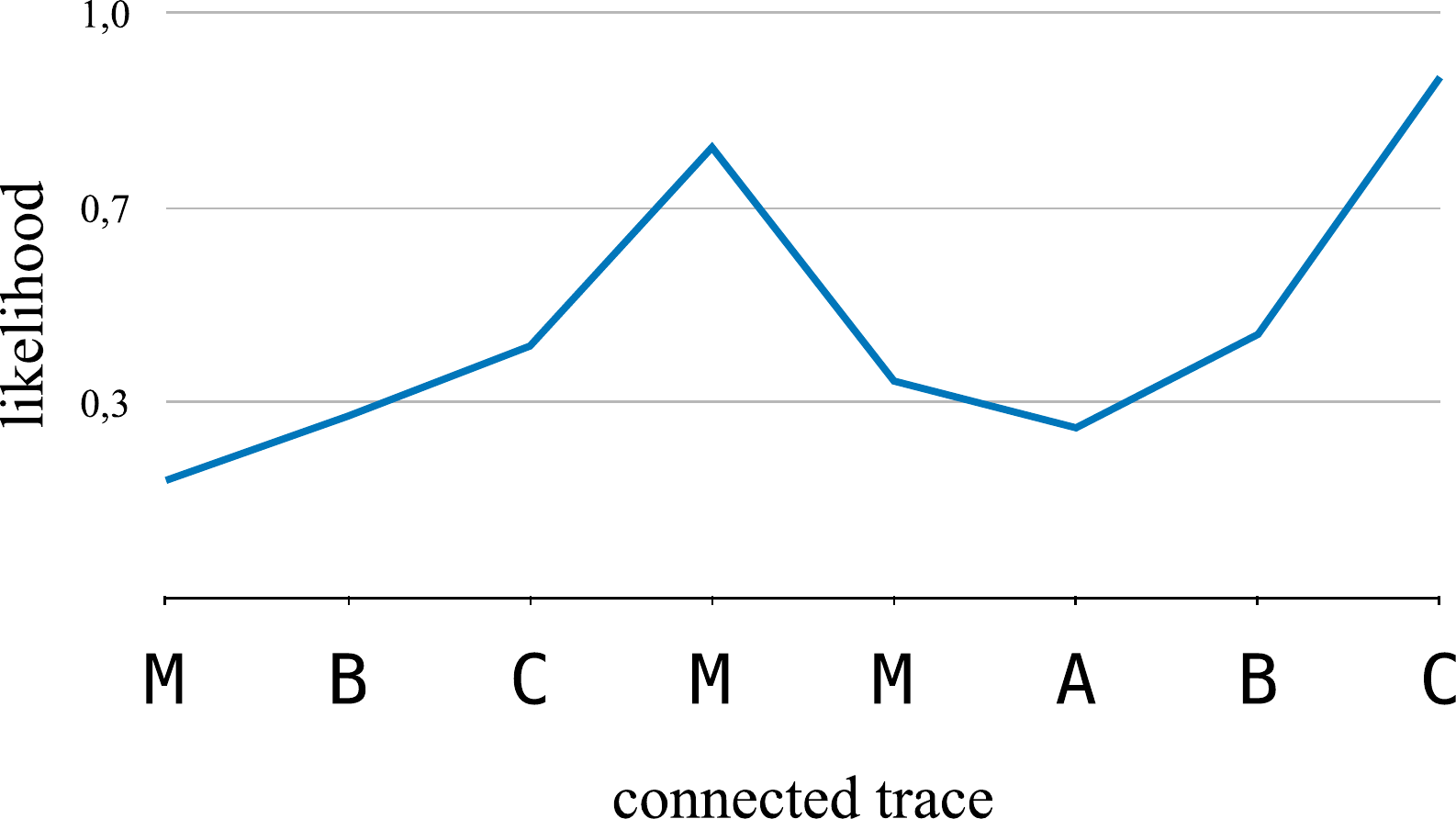}
	\caption{A plot indicating the chances of having a case segment for each position of the user interaction data (second and third trace from the example in Figure~\ref{fig:dfg}).}
	\label{fig:segment}
\end{figure}

Through the word2vec model we trained in Section~\ref{sec:training}, we can now estimate the likelihood of a case boundary $\blacksquare$ at any position of a sequence of user interactions. Figure~\ref{fig:segment} shows these estimates on one user interaction sequence from the running example. Note that this method of computing likelihoods is easy to extend to an ensemble of predictive models: the different predicted values can be then aggregated, e.g., with the mean or the median.

Next, we use these score to determine case boundaries, which will correspond to prominent peaks in the graph. Let $\langle p_1, p_2, \dots, p_n \rangle$ be the sequence of likelihoods of a case boundary obtained on a user interaction sequence. We consider $p_i$ a boundary if it satisfies the following conditions: first, $p_i > b_1 \cdot p_{i-1}$; then, $p_i > b_2 \cdot p_{i+1}$; finally, $p_i > b_3 \cdot \frac{\sum_{j=i-k-1}^{i-1} p_j}{k}$, where $b_1, b_2, b_3 \in [1, \infty)$ and $k \in \mathbb{N}$ are hyperparameters that influence the sensitivity of the segmentation. The first two inequalities use $b_1$ and $b_2$ to ensure that the score is sufficiently higher than the immediate predecessor and successor. The third inequality uses $b_3$ to make sure that the likelihood is also significantly higher than a neighborhood defined by the parameter $k$.

These three conditions allow us to select valid case boundaries within user interaction sequences. Splitting the sequences on such boundaries yields traces of complete process executions, whose events will be assigned a unique case identifier. The set of such traces then constitutes a traditional event log, ready to be analyzed with established process mining techniques.

In the following two sections, we will evaluate two important aspects of our method. Section~\ref{sec:perf} examines the time performance of the method, and verifies whether it is feasible for large user interaction logs. Section~\ref{sec:qual} validates our method qualitatively, through a user study in a real-world setting.

\section{Time performance}\label{sec:perf}

Let us now see the efficiency of our method in obtaining a segmentation model. The training phase consists in the generation of the training set and the transition system, and the training of the underlying word2vec models. These steps can take up a considerable amount of time depending on the log size and therefore have to be considered.

\begin{figure}[h]
	\centering
	\includegraphics[width=.9\textwidth]{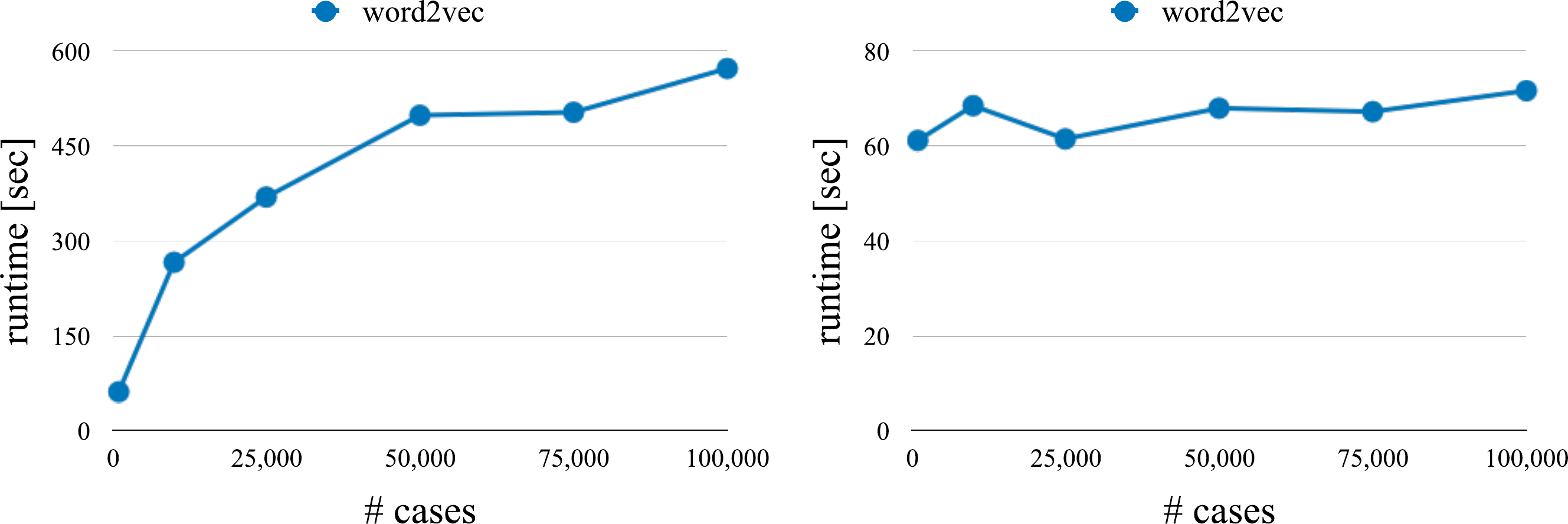}
	\caption{The runtime of the proposed method during the generation of the training log (left) and the time that is required for the model training (right) depending on the number of cases in the input log.} 
	\label{fig:time_gen_training}
\end{figure}

In Figure \ref{fig:time_gen_training} it can be seen that the time required for the generation of the training set (left) increases quickly for small to medium sized logs, but then plateaus for larger logs. The main factor for the performance of the training set generation is the complexity of the underlying transition system. A larger log will generally contain more behavior, which in turn will lead to a more complex transition system. More paths therefore have to be considered during the generation of the artificial traces. This may explain the plateauing for larger logs; beyond a certain amount of traces, increasing the size of the log will no longer significantly increase the number of variants it contains. The number of states and transitions in the transition system will therefore stop growing, since the system already depicts all of the possible behavior. After this point, the performance of the generation will plateau and is no longer depending on the size of the log.

For the training of the word2vec models, we see a constant required time with minor fluctuations. This indicates that there is no influence of the size of the training log on the performance of the model training. This is caused by the fact that the size of the artificial training log does not depend on the size of the input log, but can be freely chosen. Since the same sized training set was used for all of the logs, the training time did not change significantly.

\begin{figure}[t]
	\centering
	\includegraphics[width=0.48\textwidth]{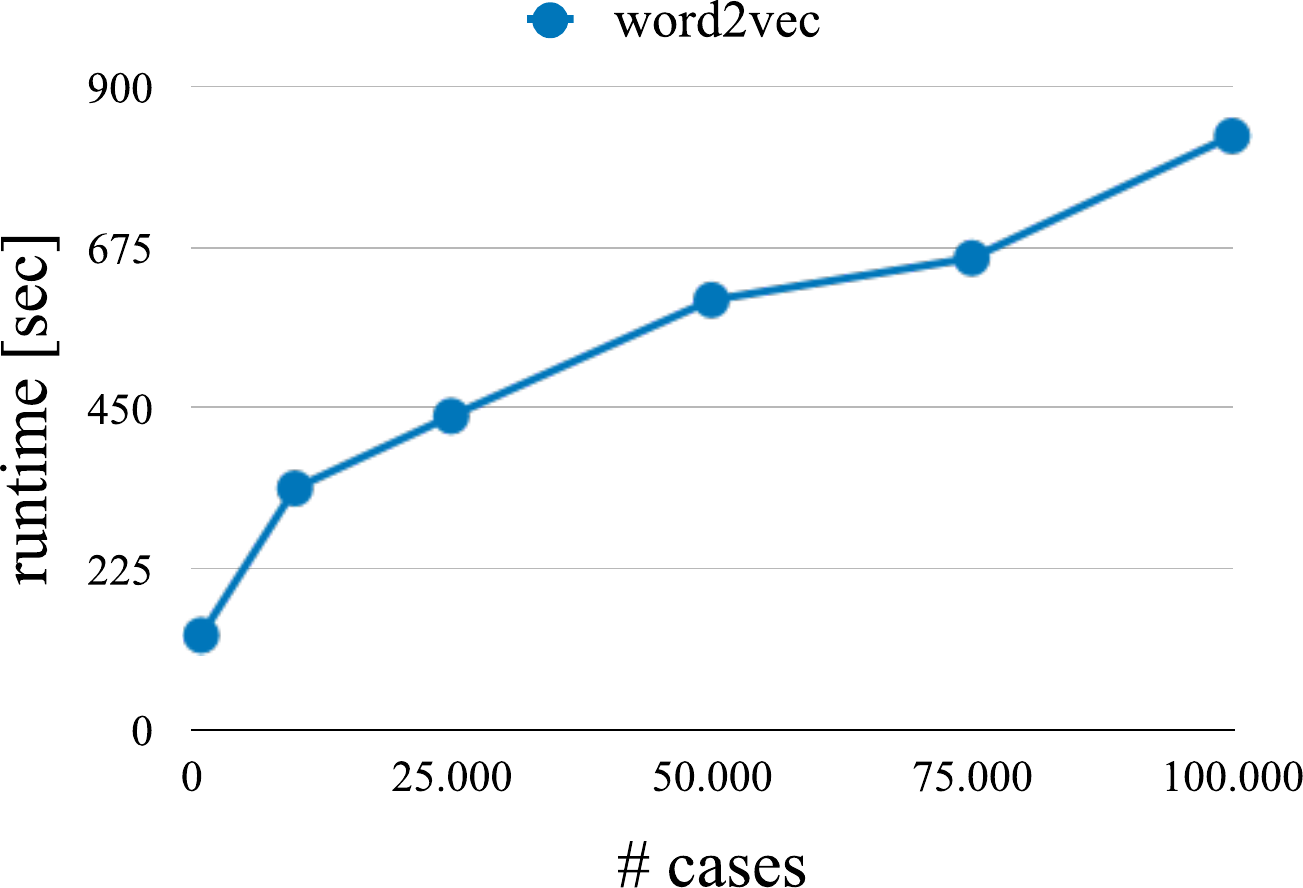}
	\caption{The overall runtime of the proposed method in the preparation phase depending on the number of cases in the input log.}
	\label{fig:time_overall}
\end{figure}

The combined time that is required for the complete preparation phase of the proposed method, depending on the size of the input log, can be seen in Figure \ref{fig:time_overall}. The overall time is mainly influenced by the generation of the transition system, since the model training requires a constant time. Other parts of the preparation phase such as the computation of the required log statistics have a linear runtime and contribute to the overall runtime behavior that can be seen in Figure \ref{fig:time_overall}.

In conclusion, the preparation phase consists of steps with a time complexity of $\mathcal{O}(T'+S'^2)$ for computing the paths in the underlying transition system $(S',A,T',i)$ and a constant time complexity (model training). The size of the transition system depends on the size of the input log, but is limited by the number of variants in the log. Overall, it can therefore be said that the time performance of the preparation phase is reasonable (approximately linear in the size of the input) even for larger interaction logs, especially considering that it only has to be performed once, but may be reused for multiple segmentations.

\section{User Study}\label{sec:qual}

In order to validate the utility of process mining workflows in the area of user behavior analysis, a user study was conducted. Such study also aims at assessing the quality of the segmentation produced by the proposed method in a real-life setting, in an area where the ground truth is not available (i.e., there are no normative well-formed cases).

\subsection{Setting and Methodology}

We applied our proposed case segmentation method to a dataset which contains real user interaction data collected from the mobile applications of a German vehicle sharing company. We then utilized the resulting segmented log to analyze user behavior with an array of process mining techniques. Then, the results were presented to process experts from the company, who utilized such results to identify critical areas of the process and suggest improvements. Since the data is from a real-life case study where there is no known ground truth on the actual behavior of the users in the process, we validate our method in a qualitative way, through an assessment by process experts that the insights obtain through process mining are sensible, truthful, and useful.

In the data, the abstraction for recorded user interactions is the screen (or page) in the app. For each interaction, the system recorded five attributes: \texttt{timestamp}, \texttt{screen}, \texttt{user}, \texttt{team}, and \texttt{os}. The \texttt{timestamp} marks the point in time when the user visited the screen, which is identified by the \texttt{screen} attribute, our activity label. The \texttt{user} attribute identifies who performed the interaction, and the \texttt{team} attribute is an additional field referring to the vehicle provider associated with the interaction. Upon filtering out pre-login screens (not associated with a \texttt{user}), the log consists of about 990,000 events originating from about 12,200 users. A snippet of these click data was shown in Table~\ref{table:data}, in Section~\ref{sec:introduction}.

\subsection{Results}

After applying the segmentation method presented in Section~\ref{sec:method} to the click data, as described in the previous section, we analyzed the resulting log with well-known process mining techniques, detailed throughout the section. The findings were presented to and discussed with four experts from the company, consisting of one UX expert, two mobile developers and one manager from a technical area. All of the participants are working directly on the application and are therefore highly familiar with it. We will report here the topics of discussion in the form of questions.\\

\textbf{Q1: What is the most frequent first screen of an interaction?}

\noindent The correct answer to this question is the \texttt{station\_based\_map\_dashboard}, which could be computed by considering the first screens for all cases that were identified by the proposed method. All of the participants were able to answer this question correctly. This is expected, as all of the participants are familiar with the application. However, the answers of the participants did not distinguish between the three different types of dashboard that exist in the app. The fact that the map based dashboard is the most frequently used type of dashboard was new and surprising for all of the participants.\\ 

\noindent \textbf{Q2: What is the most frequent last screen of an interaction?}

\noindent The answer to this question can be obtained analogously to that of Q1 directly from the segmented log. In contrast to Q1, not all participants were of the same opinion regarding the answer to this question. Two participants gave the correct answer, which again is the \texttt{station\_based\_map\_dashboard}. The other two participants chose the \texttt{booking} screen. This screen is the third most frequent case end screen following the \texttt{pre\_booking} screen. After the correct answer was revealed, one participant proposed that the users may be likely to return to the dashboard after they have completed their goal. This theory can be supported with the available data. It seems that the users have an urge to \emph{clean up} the application and return it to a neutral state before leaving it. Overall, it can be concluded that the participants have a good understanding of the frequent start and end screens of the application. However, the analysis provides more detailed information and was therefore able to discover aspects about the process that were new for the experts.\\

\noindent \textbf{Q3: What is the most frequent interaction with the app?}

\noindent This question is asking about the most frequent case variants that are contained in the given log and the associated task of the user. Since the most frequent variants will usually be the shortest variants and a case consisting of only two generic screens cannot be interpreted as a task of the user in a meaningful way, these short variants were not considered for the answer to this question. According to the segmented log, the most common interaction of this type is, selecting a vehicle on the dashboard and checking its availability from the pre-booking screen. One of the four participants did answer this question correctly. Two participants answered that searching for a vehicle on the dashboard is the most frequent interaction, which is closely related to the correct answer but does not include the availability check. The remaining participant answered, opening a booking from the list of all bookings. The results again show that the participants have a good understanding of the usage of the application, but are not able to provide details that are made visible by the log analysis.\\

\noindent \textbf{Q4: What is the average length of an interaction with the app?}

\noindent For this question, the length of an interaction describes the number of interactions that belong to a case. The correct answer is $4.8$ screens, which is rather short. The participants gave the individual answers 50, 30, 12 and 10 screens, which overall results in an average of $25.5$. We see that the participants significantly overestimate the length of an average interaction with the app according to the segmented log. However, the average case length is strongly influenced by the employed case attribution method. The mismatch between the results from the log analysis and the expert opinions could therefore be caused by the segmentation that was produced by the proposed method. However, the observed deviations regarding the number of cases were overall not larger than about 50\%, which does not explain the large difference between the experts expectations and the calculated value. In order to further examine this, the result was compared to that of a time based segmentation with a fixed threshold of five minutes. These case attribution techniques tend to overestimate the length of cases, as they are not able to distinguish between cases that happen directly after each other. For this reference segmentation, an average case length of $6.7$ was calculated. This is comparable to the result of the proposed method and confirms the observation that the experts tend to overestimate the length of interactions significantly.\\

\noindent \textbf{Q5: What is the median duration of an interaction with the app?}

\noindent For this question, the median duration is used instead of the average, as outliers that have case durations of several days are skewing the average disproportionately. According to the segmented log, the median case duration is $53.4$ seconds. The participants gave the answers 240 seconds, 120 seconds, 90 seconds and 60 seconds, leading to an overall average of 127.5 seconds. Similar to the average length of the interactions, the participants did also overestimate their median duration. Only one participant did give an answer that was close to the calculated value. Both, the significant overestimation of the interaction length and the duration, show that the experts were not able to accurately assess the time a user needs in order to complete a task. This type of analysis is not possible using an unsegmented log and was therefore enabled by the use of the proposed method.\\

\noindent \textbf{Q6: How does the median interaction duration on Android and iOS compare?}

\noindent As was introduced before, for each interaction it is recorded if it occurred in the Android or iOS application. This allows the comparison between the different applications during analysis. During the analysis it was discovered that the median interaction duration on iOS of $39.4$ seconds is significantly shorter than the $92.9$ seconds observed for the Android application. The participants were not aware of this difference, as three of the four participants thought that the interaction durations would be the same between the different operating systems and one participant thought that interactions would be shorter on Android. One of the participants argued that Android users may generally be more inclined to ``play around'' within the application, which may explain the observed difference. Regarding the analysis, the observed deviation could also be caused by differences in the implementation of the screen recording between the two apps. The produced segmentation may reflect cases originating from one of the apps more accurately than those from the other, because the same task of a user may translate to a different sequence of screens in the two apps.\\

\noindent \textbf{Q7: Given that 42\% of the users use the Android app, what percentage of interactions are from Android users?}

\noindent In general one would expect that the fraction of cases that originate from the Android app is similar to the number of users that are using this operating system. The conducted analysis does however show, that only $31\%$ of cases originate from the android app, which is significantly lower than expected. The participants did not expect this uneven distribution, which is emphasized by their answers. Two participants expected a ratio of $50\%$ and two participants answered that $60\%$ of the cases originate from the Android app. In conjunction with the results for the median interaction time that were discussed in Q6/Q7, this means that according to the computed segmentation, Android users tend to use the app longer but overall less frequently.\\

\noindent \textbf{Q8: Draw your own process model of the user interactions.}

\begin{figure}[t]
	\centering
	\includegraphics[width=.8\textwidth]{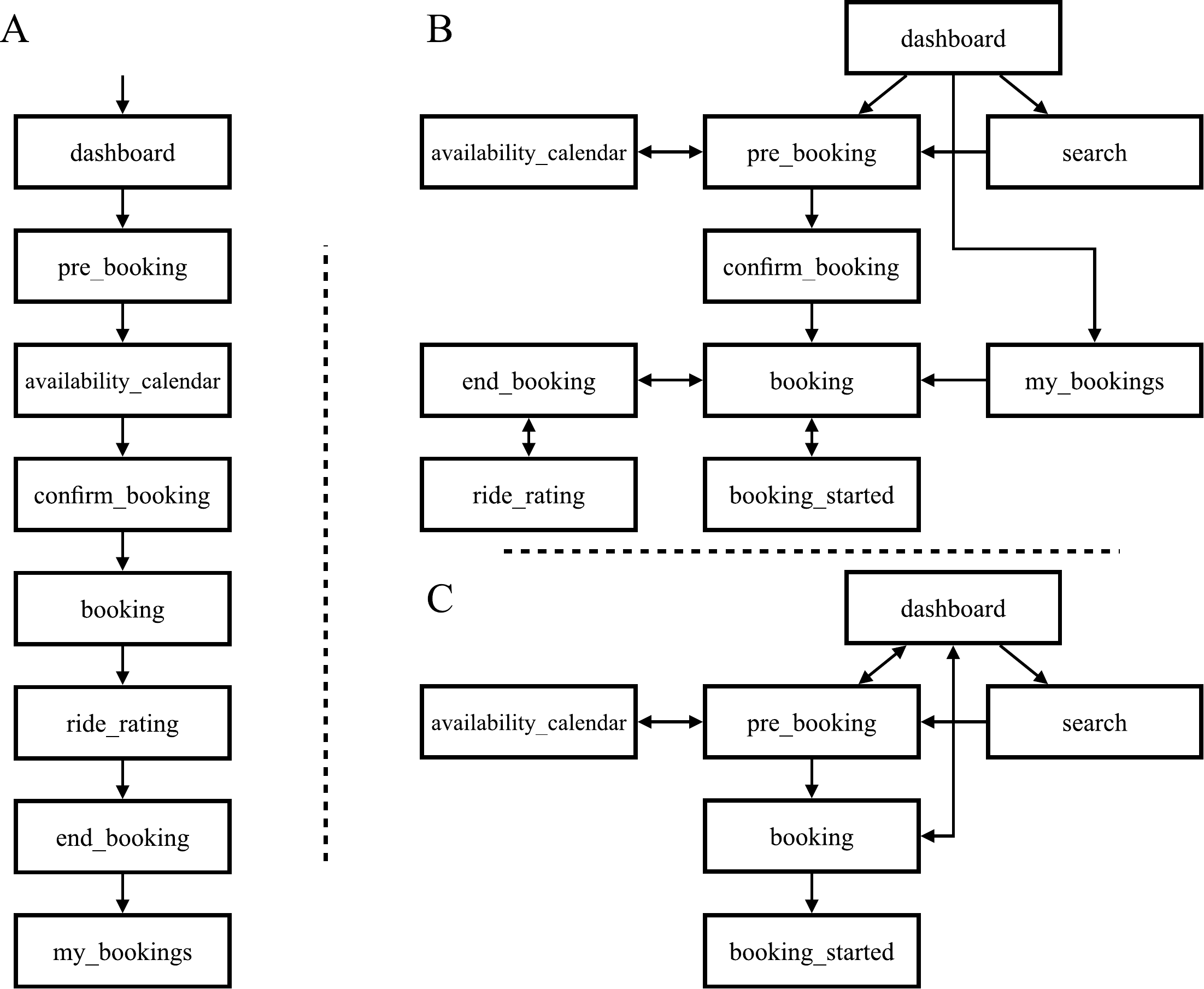}
	\caption{DFGs created by three of the process experts as part of Q1.}
	\label{fig:expert1}
\end{figure}

\begin{figure}[h!]
	\centering
	\includegraphics[width=.8\textwidth]{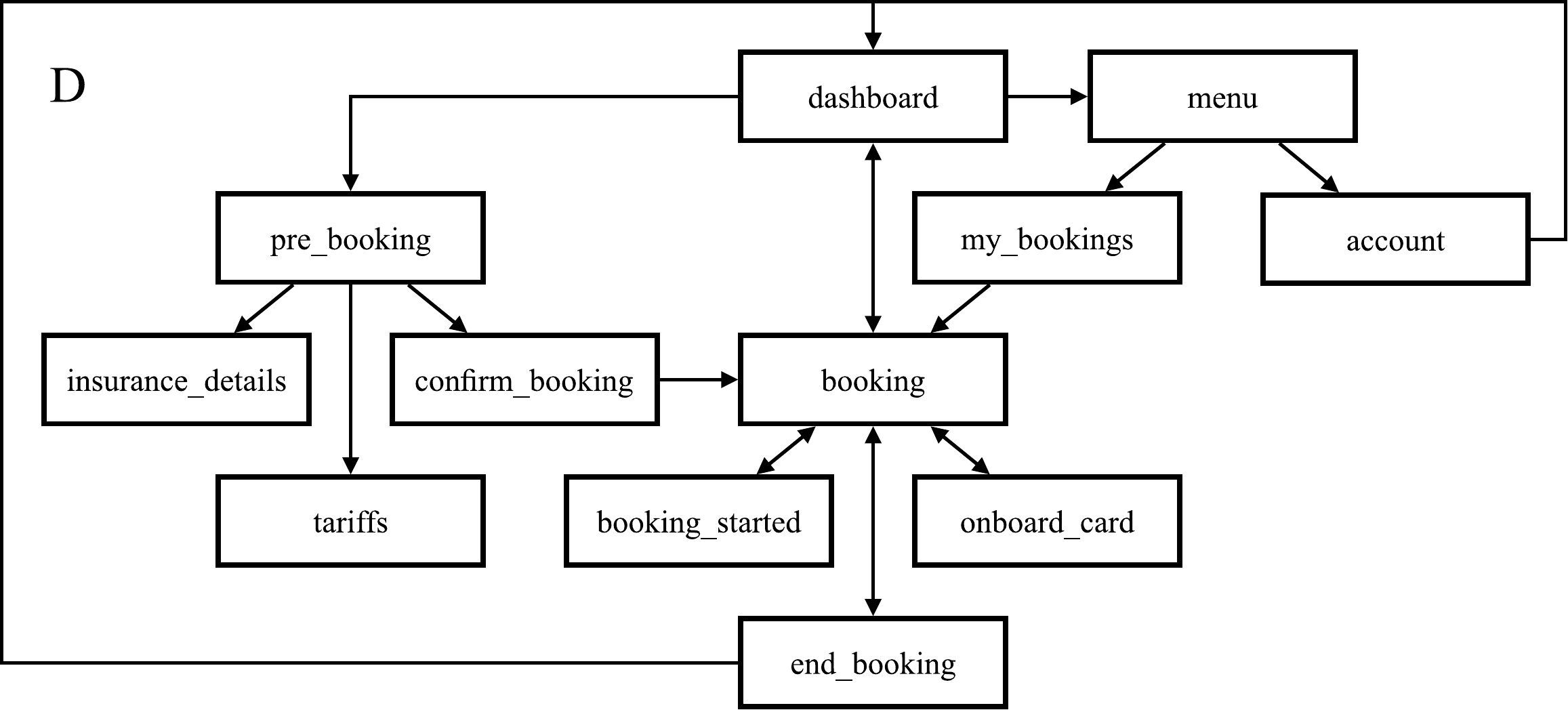}
	\caption{DFG created by one of the process experts as part of Q1.}
	\label{fig:expert2}
\end{figure}

\noindent The participants were asked to draw a \emph{Directly-Follows Graph} (DFG) describing the most common user interactions with the app. A DFG is a simple process model consisting in a graph where activities A and B are connected by an arc if B is executed immediately after A. The concept of this type of graph was explained to the participants beforehand. The experts were given five minutes in order to create their models. A cleaned up representation of the resulting models can be seen in Figures~\ref{fig:expert1} and~\ref{fig:expert2}.

\begin{figure}[t]
	\centering
	\includegraphics[width=.8\textwidth]{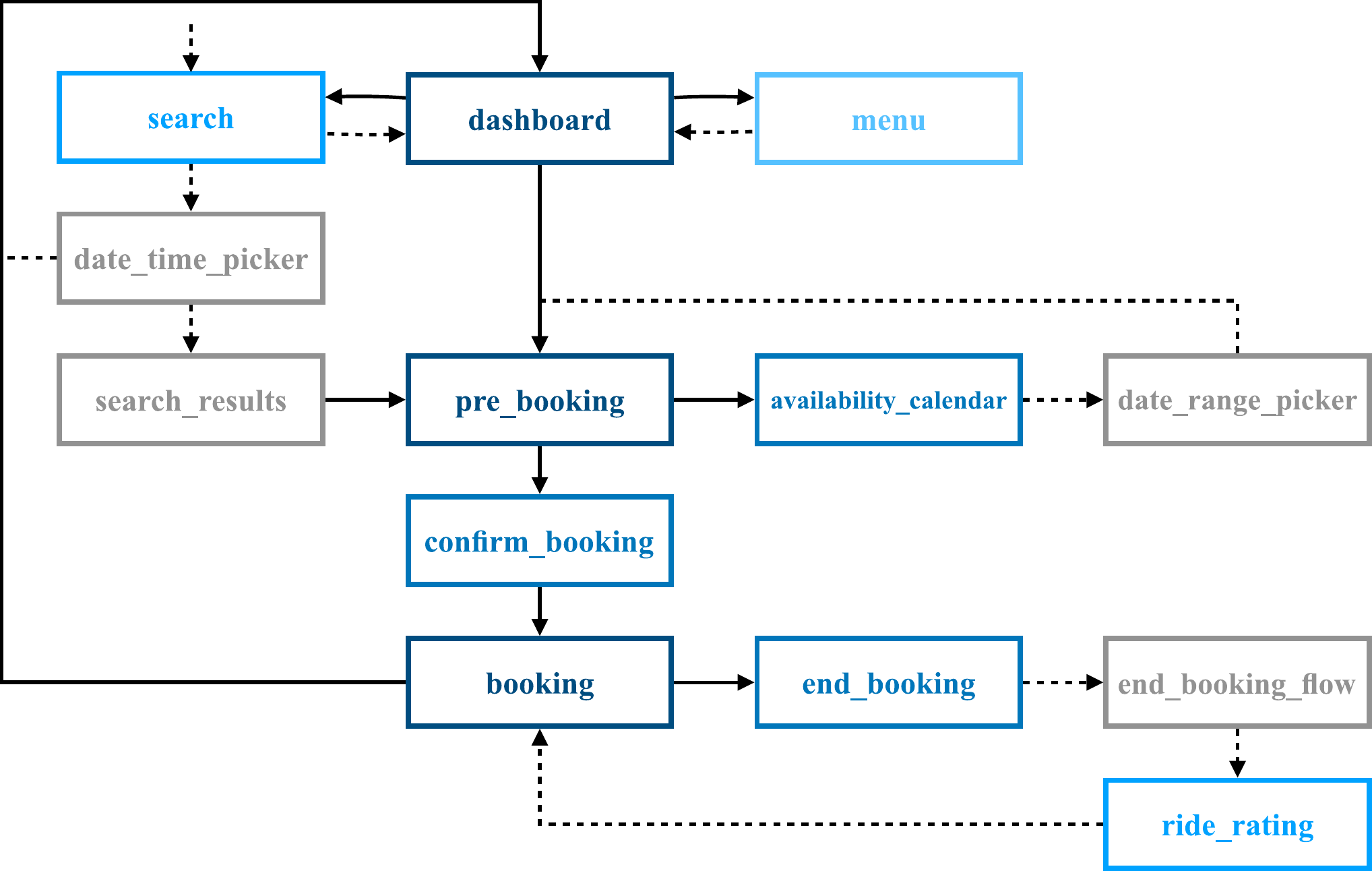}
	\caption{DFG automatically discovered from the log segmented by our method. Darker activities and solid edges were included in models hand-drawn by participants; light-colored activities and dashed edges were not identified by the majority of participants.}
	\label{fig:true}
\end{figure}

For comparison, we created a DFG of the segmented log (Figure~\ref{fig:true}). Such model was configured to contain a similar amount of different screens as the expert models. The colors indicate the agreement between the model and the expert models. Darker colors signify that a screen was included in more expert models. The dashed edges between the screens signify edges that were identified by the generated model, but are not present in the participant's models.

The mobile developers (models A and B) tend to describe the interactions in a more precise way that follows the different screens more closely, while the technical manager and UX expert (C and D) provided models that capture the usage of the application in a more abstract way. The fact that the computed model and the expert models are overall very similar to each other suggests that our proposed method is able to create a segmentation that contains cases that are able to accurately describe the real user behavior.\\

\noindent \textbf{Q9: Given this process model that is based on interactions ending on the \texttt{booking} screen, what are your observations?}

\begin{figure}[t]
	\centering
	\includegraphics[width=.8\textwidth]{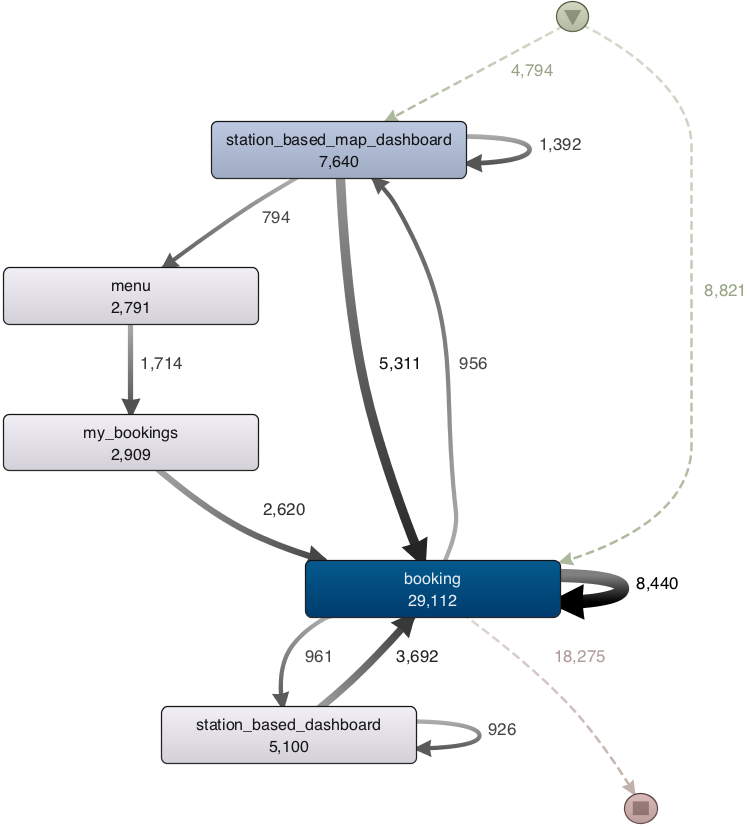}
	\caption{A process model created by using Disco~\cite{DBLP:conf/bpm/GuntherR12}, with the \texttt{booking} screen as endpoint of the process.}
	\label{fig:discofreq}
\end{figure}

\noindent Given the process model shown in Figure~\ref{fig:discofreq}, the participants were surprised by the fact that the map-based dashboard type is used significantly more frequently than the basic dashboard is surprising to them. Additionally, two of the experts were surprised by the number of users that are accessing their bookings through the list of all bookings (\texttt{my\_bookings}). This latter observation was also made during the analysis of the segmented log and is the reason that this process model was presented to the experts. In general, a user that has created a booking for a vehicle can access this booking directly from all of the different types of dashboards. The fact that a large fraction of the users takes a detour through the menu and booking list in order to reach the booking screen is therefore surprising. This circumstance was actually already identified by one of the mobile developers some time before this evaluation, while they were manually analyzing the raw interaction recordings data. They noticed this behavior because they repeatedly encountered the underlying pattern while working with the data for other unrelated reasons. Using the segmented user interaction log, the behavior was however much more discoverable and supported by concrete data rather than just a vague feeling. Another observation that was not made by the participants is that the path through the booking list is more frequently taken by users that originate from the map-based dashboard rather than the basic dashboard. The UX expert suspected that this may have been the case, because the card that can be used to access a booking from the dashboard is significantly smaller on the map-based dashboard and may therefore be missed more frequently by the users. This is a concrete actionable finding of the analysis that was only made possible by the use of process mining techniques in conjunction with the proposed method.\\

\noindent \textbf{Q10: Given this process model that is based on interactions ending on the \texttt{search} screen, what are your observations?}

\begin{figure}[h!]
	\centering
	\includegraphics[width=\textwidth]{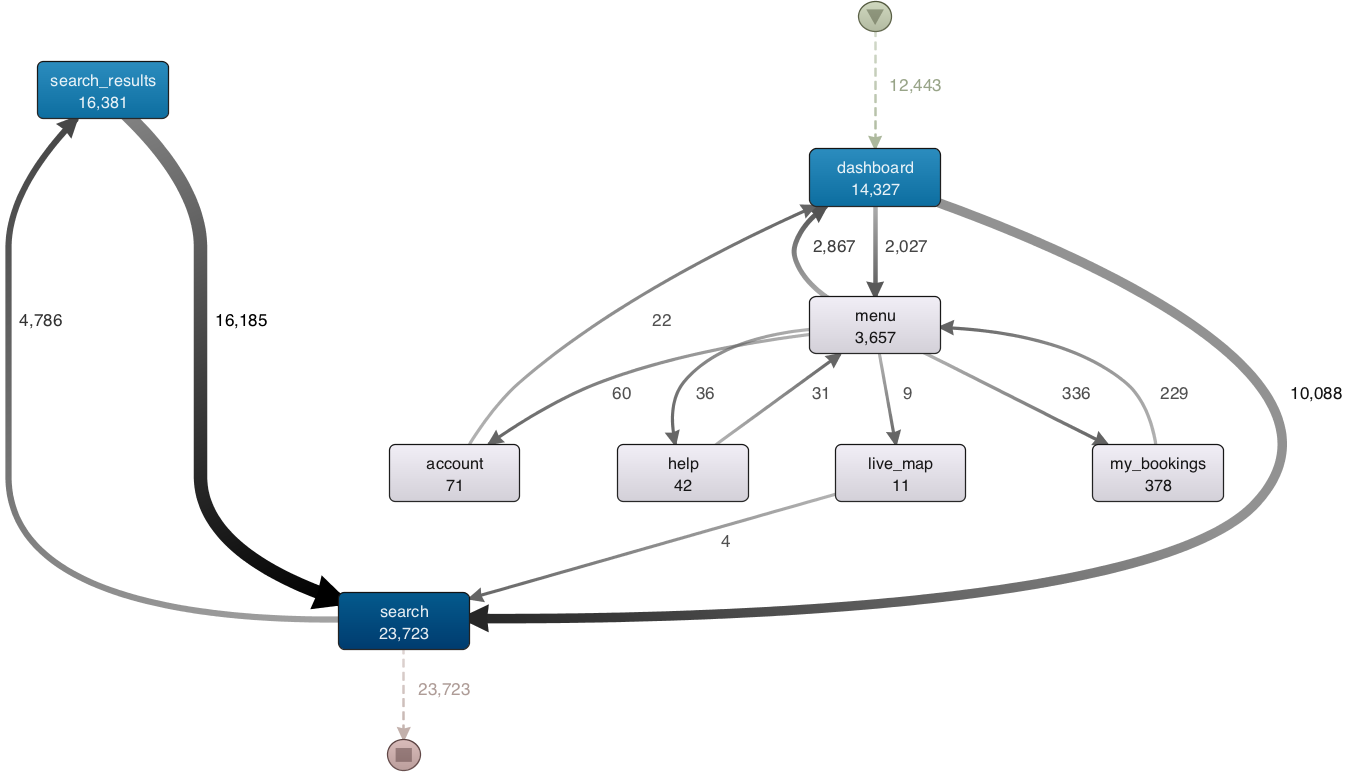}
	\caption{A process model with the \texttt{search} screen as endpoint of the process.}
	\label{fig:model_q11}
\end{figure}

The behavior that was observed during the analysis was tried to be conveyed to the participants using the model that can be found in Figure \ref{fig:model_q11}. Since the model is based on all cases including the \texttt{search} screen, which start at any type of dashboard, and the \texttt{search} screen is directly reachable from the dashboards, it would be expected that no significant amount of other screens are included in the model. This is however not the case, as the \texttt{menu} screen and the various screens that are reachable from this screen are included in many of the cases that eventually lead to a search. This suggests that the users that did want to perform a search, tried to find the \texttt{search} screen in the main menu, implying that it is not presented prominently enough on the dashboards. None of the experts had this observation when they were presented the discussed model.\\

\noindent \textbf{Q11: What is the median time a user takes to book a vehicle?}

\noindent The correct answer to this question is 66 seconds. This was calculated based on the median time of all cases in which a vehicle booking was confirmed. Three participants gave the answers 420 seconds, 120 seconds and 120 seconds. The fourth participants argued that this time may depend on the type of dashboard that the user is using and answered 300 seconds for the basic dashboard and 120 seconds for the map-based dashboard. When asked to settle on only one time, the participant gave an answer of 180 seconds. Overall this means that the experts estimated a median duration for this task of 3 minutes and 30 seconds. This again is a significant overestimation compared to the value that was obtained by analyzing the real user behavior. Again, a mismatch between the perception of the experts and the real behavior of the users was revealed.\\

\begin{figure}[]
	\centering
	\includegraphics[width=\textwidth]{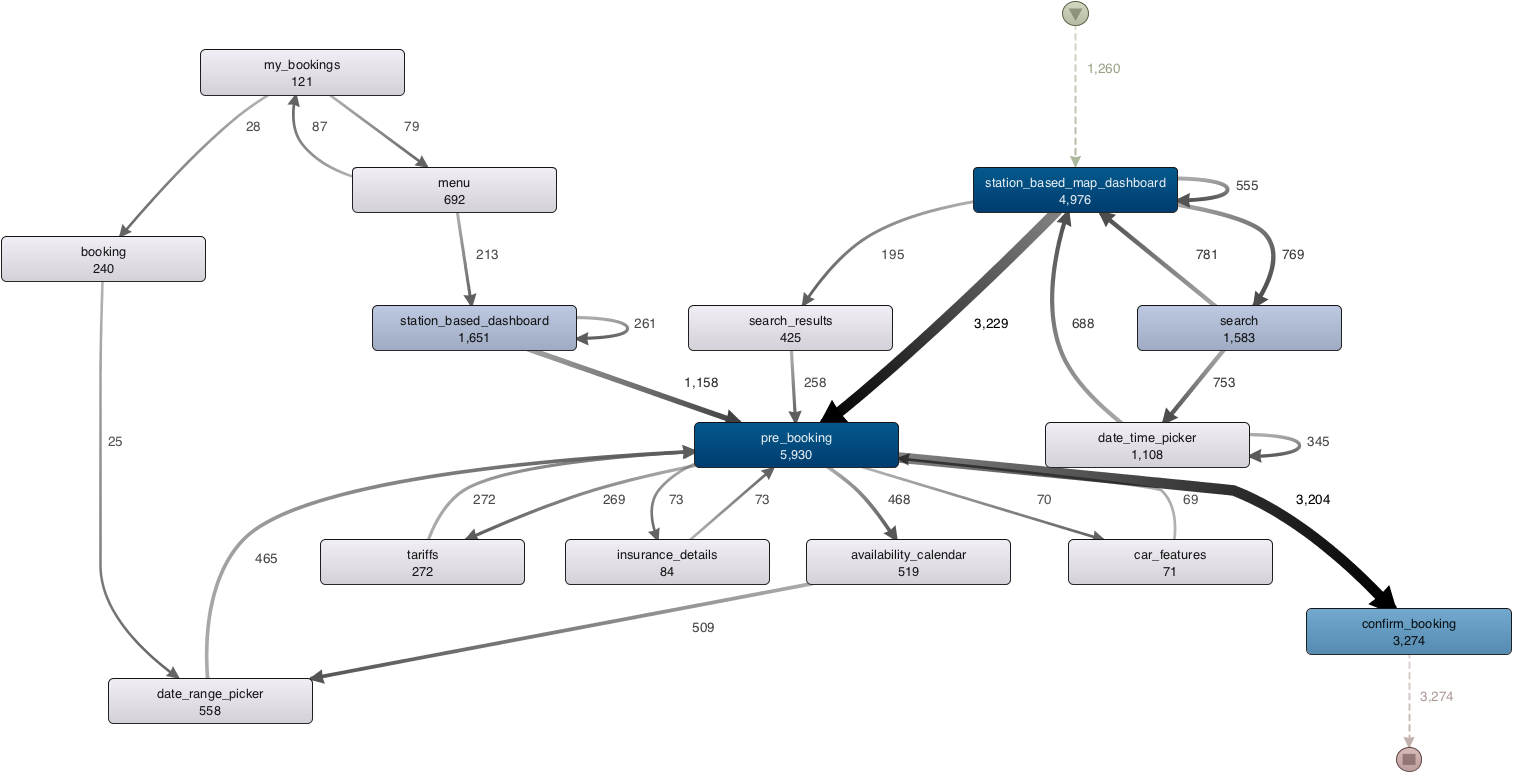}
	\caption{A process model based on cases that begin in any dashboard and end on the \texttt{confirm\_booking} screen.}
	\label{fig:modelconfirm}
\end{figure}

\noindent \textbf{Q12: Given this process model that is based on interactions ending on the \texttt{confirm booking} screen (Figure~\ref{fig:modelconfirm}), what are your observations?}

\noindent Several of the experts observed that the screens that show details about the vehicles and the service, such as \texttt{tariffs}, \texttt{insurance\_details} and \texttt{car\_features}, are seemingly used much less frequently than expected. In only about 2-10\% of cases, the user visits these screens before booking a vehicle. When considering the concrete numbers, the \texttt{availability\_calendar} screen (which is used to choose a timeframe for the booking) and the \texttt{tariffs} screen (which displays pricing information) are used most frequently before a booking confirmation. This suggests that time and pricing information are significantly more important to the users than information about the vehicle or about the included insurance. These findings sparked a detailed discussion between the experts about the possible reasons for the observed behavior. Nonetheless, this shows that models obtained from segmented user interaction logs are an important tool for the analysis of user behavior and that these models provide a valuable foundation for a more detailed analysis by the process experts. Another observation regarding this model was, that a majority of the users seem to choose a vehicle directly from the dashboard cards present on the app rather than using the search functionality. This suggests that the users are more interested in the vehicle itself, rather than looking for any available vehicle at a certain point in time.\\

\noindent \textbf{Q13: Discuss the fact that 2\% of users activate the intermediate lock before ending the booking.}

\noindent The smartphone application offers the functionality to lock certain kinds of vehicles during an active booking. This is for example possible for bicycles, which can be locked by the users during the booking whenever they are leaving the bicycle alone. To do so, the \texttt{intermediate\_lock} and \texttt{intermediate\_action} screens are used. During the analysis, it was found that 2\% of users use this functionality in order to lock the vehicle directly before ending the booking. This is noteworthy, as it is not necessary to manually lock the vehicle before returning it. All vehicles are automatically locked by the system at the end of each booking. One expert argued that this may introduce additional technical difficulties during the vehicle return, because the system will try to lock the vehicle again. These redundant lock operations, discovered analyzing the segmented log, may introduce errors in the return process.\\

\noindent \textbf{Q14: Discuss the fact that only 5\% of users visit \texttt{damages} and \texttt{cleanliness}.}

\noindent The application allows users to report damages to the vehicles and rate their cleanliness, through the homonymous pages. It was possible to observe that only a small percentage of the users seem to follow this routine, which was surprising to the experts. For the vehicle providers it is generally important that the users are reporting problems with the vehicles; optimally, every user should do this for all of their bookings. According to the data, this is however not the case, as only a small percentage of the users are actually using both of the functionalities. The experts, therefore, concluded that a better communication of these functionalities is required.

\subsection{Discussion}

In this section, we will consider and discuss some aspects, advantages, and limitations of our approach and its applications

\begin{figure}[t]
	\centering
	\includegraphics[width=.89\textwidth]{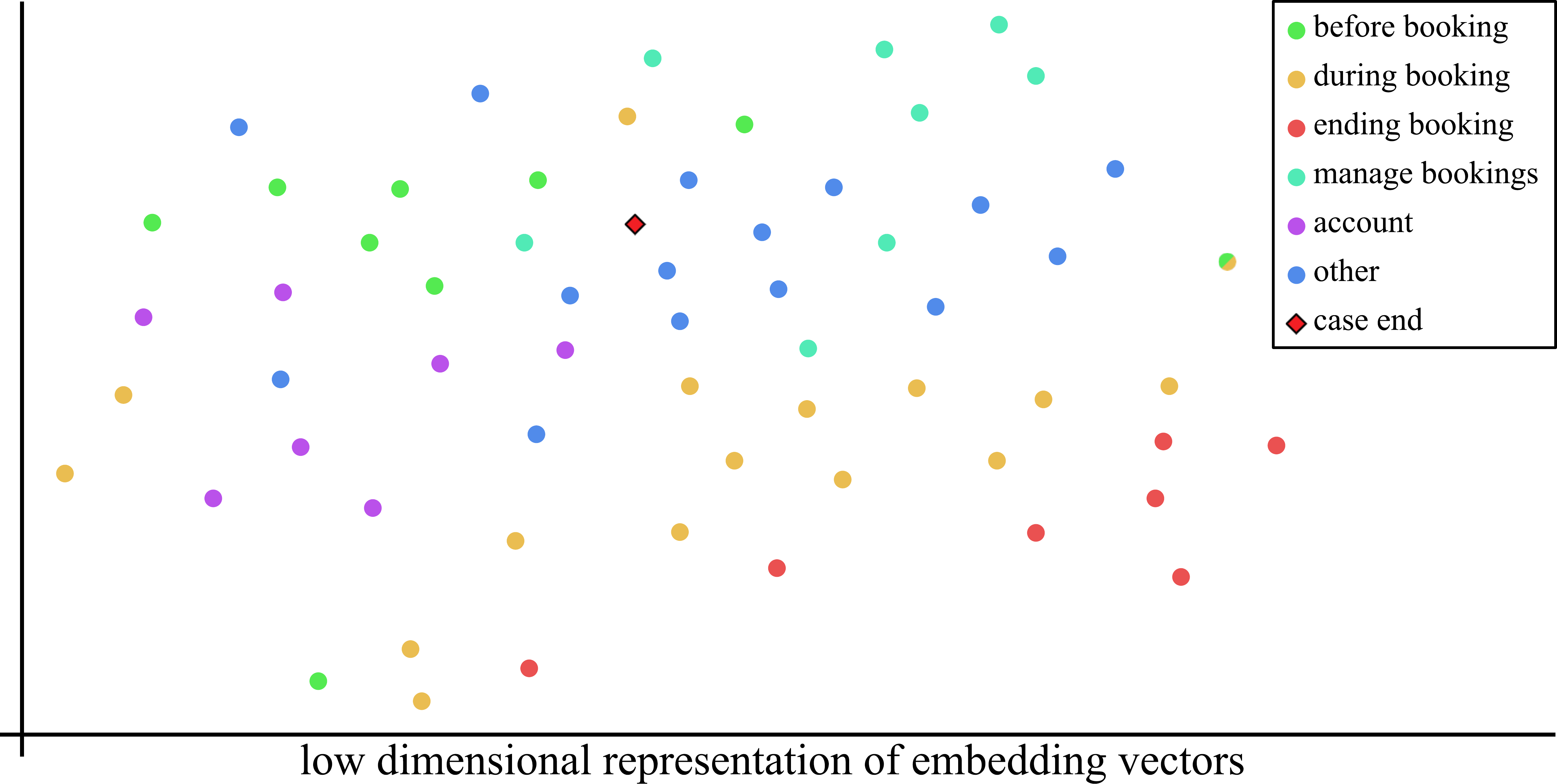}
	\caption{A two dimensional representation of the activity embedding vectors of a word2vec model that was trained in the context of the case study. Each dot represents the relative location of an action embedding. The closer two dots are, the more similar are their corresponding embedding vectors. The different colors represent different phases of the process; we can see that similarly colored activities tend to form clusters in the vector space. The dimensional reduction is based on the t-SNE method \cite{van2008visualizing}.}
	\label{fig:low_dim_vectors}
\end{figure}

In order to evaluate how well the proposed method is able to capture the behavior in the input log and the semantic relationships between activities, we will visualize the embedding vectors of the trained word2vec model. Figure \ref{fig:low_dim_vectors} depicts a low dimensional representation of these embedding vectors. The model was trained with the interaction log that was the basis for the conducted case study. The different colors of the dots indicate the different areas of the application. When two actions (dots) are closer to each other in this representation, the actions are related and occur in similar contexts according to the trained model.

Activities that occur during the same phase of the usage will be close to each other in the vector space, and will form clusters. Such clustering of different kinds of actions can be observed in Figure \ref{fig:low_dim_vectors}. We can see that similar activities indeed form clusters; especially noticeable are the clusters of actions belonging to more distinct phases of the process, such as actions that occur before, during, or at the end of a booking. It can also be observed that the clusters of phases that are more similar to each other are closer to each other in the diagram. For example, the cluster of actions that occur before the booking are closer to those actions that happen during the booking and farther from the ones at the end of the booking. The overall flow of a common interaction with the application is recognizable in the diagram. This recognizable structure in the activity embedding vectors suggests that the underlying word2vec models is able to abstract the underlying process.

The embedding of the artificial end action that is introduced before model training is marked in Figure \ref{fig:low_dim_vectors} with a red rhombus. We can see that it is located near the center of the graphic and shows no clear bias toward any phase of the process. This however also means that end action embedding has no clear relation to any of the clusters. This is expected, as case ends may occur in all of the different phases of the process; however, this can also be considered a weak point in our method, since it indicates that the case end has limited specificity with respect to the type of other activities. One possible solution to this problem that would make the end action more specific is to introduce multiple different end actions, depending on the different process phases, through either different data pre-processing or a post-processing phase on the resulting embeddings.

Even though we applied basic and easily-interpretable process mining techniques to the resulting segmented event log, our user study shows the potential of the application of process mining to user behavior analytics. It was made clear by the study that the process experts are able to comprehend the basic structure of the application and therefore the underlying process well. However, whenever a more detailed view of one aspect of the process was considered, the experts were not able to correctly and accurately assess the real behavior of the users. For instance, concerning the modeling of the process, the experts were able to identify the structure of the most common interactions, but lacked detail and accuracy. This is especially true when considering the transitions between different screens. The automatically discovered model was more comprehensive and included more behavior and detail.

When the analysis results are processed, visualized and presented to the experts in the right way, they were able to produce clear and actionable results based on the findings. For example it was shown that interactions with the app are much shorter than predicted, that the users are utilizing the bookings list much more frequently than expected, that the map dashboard is the most frequently used dashboard, that the search is less important than the dashboard suggestion cards or that users are unnecessarily locking their vehicles before returning them. Based on these findings, the experts are able to derive concrete and actionable changes to the application, with the goal of improving the overall user experience. Many of the results were completely new and unexpected to the experts and were only enabled through the use of the real dataset in conjunction with the proposed case attribution approach. The time that was required for the segmentation of the large provided interaction log and the subsequent analysis is negligible compared to the amount of information that was obtained.

Overall, the experts were impressed by the findings of the analysis and were able to obtain new insights into the way their users are using the application that were not possible before. Concrete suggestions for improvements could be made and will in the future be implemented in order to improve the user experience of the application, in turn improving the customer satisfaction and lower the required support effort.

\section{Related Work}\label{sec:related}

\subsection{Event-Case Correlation}

The problem of assigning a case identifier to events in a log is a long-standing challenge in the process mining community~\cite{DBLP:conf/bpm/FerreiraG09}, and is known by multiple names in literature, including \emph{event-case correlation} problem~\cite{DBLP:conf/er/BayomieCRM19} and \emph{case notion discovery} problem~\cite{DBLP:journals/kais/MurillasRA20}. Event logs where events are missing the case identifier attribute are usually referred to as \emph{unlabeled event logs}~\cite{DBLP:conf/bpm/FerreiraG09}.

The lack of a case notion has been identified as a major challenge in a number of practical applications, such as analyzing the user interaction with the interface of CT scanners in clinical contexts~\cite{reindler2020siemens} or measuring the learnability of software systems~\cite{DBLP:conf/ACMdis/MarrellaC18}. Several of the early attempts to solve this problem, such as an early one by Ferreira and Gillblad based on first order Markov models~\cite{DBLP:conf/bpm/FerreiraG09}, a later approach by Ferreira et al. based on partitioning sequences such that they are minimal and represent a possible process instance~\cite{DBLP:journals/dke/WalickiF11}, or the more recent \emph{Correlation Miner} by Pourmiza et al., based on quadratic programming~\cite{DBLP:journals/ijcis/PourmirzaDG17} are very limited in the presence of loops in the process. Other approaches, such as the one by Bayomie et al.~\cite{DBLP:conf/caise/BayomieAE16} can indeed work in the presence of loops, by relying on heuristics based on activities duration which lead to a set of candidate segmented logs. This comes at the cost of a slow computing time. An improvement of the aforementioned method~\cite{DBLP:conf/er/BayomieCRM19} employs simulated annealing to select an optimal case notion; while still computationally heavy, this method delivers high-quality case attribution results. This was further improved in~\cite{DBLP:journals/corr/abs-2206-10009}, where the authors reduce the dependence of the method from control flow information and exploit user defined rules to obtain a higher quality result. It is of course important to remember that such methods solve a different and more general problem (the information about the resource is not necessary available) than the one examined in this paper; in this work, we focus in a more specific setting, where stronger assumptions hold. Such assumptions allow for more efficient segmentation methods, such as the one presented here.

A quite large family of methods approach the problem with a radically different assumption: the hypothesis is that the case information is indeed present in the log, but is hidden. In this context, the case identifier is disguised as a different attribute, or result of a combination of attributes, or learned by applying a similarity function between events. Several such approaches require user-defined rules or domain knowledge to uncover attribute correlations~\cite{DBLP:journals/vldb/Motahari-NezhadSCB11,DBLP:conf/hicss/EngelB14,DBLP:journals/isem/EngelKZPBAWH16} or require the case notion to be recognizable from a pattern search within the data~\cite{DBLP:conf/caise/AndaloussiBW18,DBLP:conf/ifip8-1/BalaMSQ18}.

Many available UI logs are obtained by tracking user action throughout the use of an application, software, or other systems. This means that, similarly to the case study of this paper---which contains roughly one million events---interaction logs are often of large dimensions, at least compared to the typical log sizes in process mining. Therefore, efficiency is important, especially at scale. This motivated our design of a novel method able to reconstruct a case notion for the special case user interaction logs in a fast, interpretable, and loop-robust way, and without relying on ground truth information on cases. This work is an extended version of previous results~\cite{DBLP:conf/bpmds/0001UHA22}; we hereby integrate our paper with a more formal description of the method, an evaluation on the time performance of our log segmentation approach, and a full reportage on our mobility app process mining user study.

\subsection{Uncertain Event Data}

The problem of event-case correlation can be positioned in the broader context of \emph{uncertain event data}~\cite{pegoraro2021process,DBLP:conf/apn/PegoraroUA21}. This research direction aims to analyze event data with imprecise attributes, where single traces might correspond to an array of possible real-life scenarios. For instance, a given event in a log might lack the value of a discrete event attribute such as the activity label, but we might know a set of potential labels; for continuous attributes such as a timestamp, we might have an interval of possible values at our disposal. This type of meta-information on attributes can be quantified with probabilities (\emph{probabilistic uncertainty}) or not (\emph{non-deterministic uncertainty}). Akin to the method proposed in this paper, some techniques allow to obtain probability distributions over such scenarios~\cite{DBLP:conf/icpm/0001BUA21}.

Unlabeled logs can then be seen as a specific case of uncertain event logs, where the case identifier is uncertain---since it is not known. Note that having uncertain case identifiers entails more severe consequences than other known types of uncertainty: in all other types, the concept of trace is preserved. According to uncertain event data taxonomies, a missing case identifier can be seen as a stronger type of \emph{event indetermination}~\cite{pegoraro2021process}, which occurs when an event has been recorded in the log, but it is unclear if it actually happened in reality. Event indetermination is a weaker loss of information then a missing case identifier, in the sense that more information is present and some process mining techniques, albeit specialized, are still possible.

\subsection{Robotic Process Automation}

A notable and rapidly-growing field where the problem of event-case correlation is crucial is \emph{Robotic Process Automation} (RPA), the automation of process activities through the identification of repeated routines in user interactions with software systems~\cite{DBLP:books/sp/22/DumasRLPM22}. Such routines are automatically discovered from pre-processed user interaction data, then the automatability of such routines is estimated and defined, and software bots are then created to aid the users in repetitive tasks within the process, such as data field completion. As a consequence, the entire discipline of RPA is based on the availability and quality of user interaction logs, which should have a clear and defined case notion. In fact, the problem of case reconstruction is known in the field, and has been identified as a central criticality in automated RPA learning~\cite{DBLP:conf/otm/GaoZLA19} and automated RPA testing~\cite{DBLP:conf/icse/Chacon-MonteroJ19}.

Similarly to many approaches related to the problem at large, existing approaches to event-case correlation in the RPA field often heavily rely on unique start and end events in order to segment the log, either explicitly or implicitly~\cite{DBLP:conf/gi/LinnZW18,DBLP:conf/caise/RamirezR0V19,DBLP:conf/icpm/LenoADRMP20}.

\subsection{Event-Case Correlation Applications}

The problem of event-case attribution is different when considered on click data---particularly from mobile apps. Normally, the goal is to learn a function that receives an event as an independent variable and produces a case identifier as an output. In the scenario studied in this paper, however, the user is tracked by the open session in the app during the interaction, and recorded events with different user identifier cannot belong to the same process case. The goal is then to subdivide the sequence of interactions from one user into one or more sessions (cases). While in this user study we assume a prior knowledge of the app where the user interaction is recorded---the link graph---, other ad-hoc techniques to obtain a case notion or segmentation are based on different prior knowledge and different assumptions.

Marrella et al.~\cite{DBLP:conf/ACMdis/MarrellaC18} examined the challenge of obtaining case identifiers for unsegmented user interaction logs in the context of learnability of software systems, by segmenting event sequences with a predefined set of start and end activities as normative information. They find that this approach cannot discover all types of cases, which limits its flexibility and applicability. Jlailaty et al.~\cite{DBLP:conf/IEEEscc/JlailatyGB17} encounter the segmentation problem in the context of email logs. They segment cases by designing an ad-hoc metric that combines event attributes such as timestamp, sender, and receiver. Their results however show that this method is eluded by edge cases. Other prominent sources of sequential event data without case attribution are IoT sensors: Janssen et al.~\cite{DBLP:conf/icpm/JanssenMKZ20} address the problem of obtaining process cases from sequential sensor event data by splitting the long traces according to an application-dependent fixed length, to find the optimal sub-trace length such that, after splitting, each case contains only a single activity. One major limitation of this approach that the authors mention is the use of only a single constant length for all of the different activities, which may have varying lengths. More recently, Burattin et al.~\cite{DBLP:journals/dke/BurattinKNW19} tackled a segmentation problem for user interactions with a modeling software; in their approach, the segmentation is obtained exploiting eye tracking data, which allows to effectively detect the end of the user interaction with the system.

\section{Conclusion}\label{sec:conclusion}
In this paper, we showed a case and user study on the topic of the problem of event-case correlation, and presented this problem in the specific application domain of user interaction data.

We examined a case study, the analysis of click data from a mobility sharing smartphone application. To perform log segmentation, we proposed an original technique based on the word2vec neural network architecture, which can obtain case identification for an unlabeled user interaction log on the sole basis of a link graph of the system as normative information. We then presented a user study, where experts of the process were confronted with insights obtained by applying process mining techniques to the log segmented using our method. The interviews with experts confirm that our technique helped to uncover hidden characteristics of the process, including inefficiencies and anomalies unknown to the domain knowledge of the business owners. Importantly, the analyses yielded actionable suggestions for UI/UX improvements, some of which were readily incorporated in the mobile app. This substantiates the scientific value of event-log correlation techniques for user interaction data, and shows the direct benefits of the application of process analysis techniques to data from the user interaction domain. Furthermore, the user study demonstrates the validity of the segmentation method presented in this paper, and its ability of producing a coherent case notion via the segmentation of user interaction sequences. Quantitative experiments with logs of increasing size show the scalability of our method, which is able to preserve its time performance with logs of large dimensions. Lastly, we highlighted how the use of a word2vec model results in a fixed-length representation for activities which expresses some of the semantic relationships between the respective activity labels.

As future work, we intend to further validate our technique by lifting it from the scope of a user study by means of a quantitative evaluation on its efficacy, to complement the qualitative one showed in this paper. Since our segmentation technique has several points of improvement, including the relatively high number of hyperparameters, it would benefit from a heuristic procedure to determine the (starting) value for such hyperparameters. It is also possible to apply different encoding techniques for embeddings in place of word2vec, which may results in a better segmentation quality for specific interaction logs. Finally, other future work may consider additional event data perspectives, such as adding the data perspective to our technique by encoding additional attributes in the training set of the neural network model.

\bibliographystyle{splncs04}
\bibliography{bibliography}

\end{document}